\documentclass[%
 reprint,
superscriptaddress,
preprintnumbers,
nofootinbib,
 amsmath,amssymb,
 aps,
]{revtex4-1}

\usepackage{graphicx}
\usepackage{dcolumn}
\usepackage{bm}
\usepackage[colorlinks=true, linkcolor=red, citecolor=blue, urlcolor=blue]{hyperref}
\usepackage{lipsum}
\usepackage{float}
\usepackage[dvipsnames, usenames]{xcolor}
\usepackage{soul}
\usepackage{makecell}

\usepackage{float} 
\usepackage{booktabs} 
\usepackage[capitalise]{cleveref}

\allowdisplaybreaks

\begin{document}


\title{Composite asymmetric dark matter with a dark photon portal: Multimessenger tests}

\author{Saikat Das}
\email{saikatdas@ufl.edu (corresponding author)}
\affiliation{Center for Gravitational Physics and Quantum Information, Yukawa Institute for Theoretical Physics, Kyoto University, Kyoto 606-8502, Japan}
\affiliation{Department of Physics, University of Florida, Gainesville, FL 32611, USA}

\author{Ayuki Kamada}
\email{akamada@fuw.edu.pl}
\affiliation{Institute of Theoretical Physics, Faculty of Physics, University of Warsaw, ul. Pasteura 5, PL-02-093 Warsaw, Poland}

\author{Takumi Kuwahara}
\email{kuwahara@pku.edu.cn (corresponding author)}
\affiliation{Center for High Energy Physics, Peking University, Beijing 100871, China}

\author{Kohta Murase}
\email{murase@psu.edu}
\affiliation{Department of Physics; Department of Astronomy \& Astrophysics; Center for Multimessenger Astrophysics, Institute for Gravitation and the Cosmos, The Pennsylvania State University, University Park, Pennsylvania 16802, USA}
\affiliation{Center for Gravitational Physics and Quantum Information, Yukawa Institute for Theoretical Physics, Kyoto University, Kyoto 606-8502, Japan}

\author{Deheng Song}
\email{songdeheng@yukawa.kyoto-u.ac.jp (corresponding author)}
\affiliation{Center for Gravitational Physics and Quantum Information, Yukawa Institute for Theoretical Physics, Kyoto University, Kyoto 606-8502, Japan}

\date{\today}

\begin{abstract}
\noindent 
Composite asymmetric dark matter (ADM) is the framework that naturally explains the coincidence of the baryon density and the dark matter density of the Universe. 
Through a portal interaction sharing particle-antiparticle asymmetries in the Standard Model and dark sectors, dark matter particles, which are dark-sector counterparts of baryons, can decay into antineutrinos and dark-sector counterparts of mesons (dark mesons) or dark photon. Subsequent cascade decay of the dark mesons and the dark photon can also provide electromagnetic fluxes at late times of the Universe. 
The cosmic-ray constraints on the decaying dark matter with the mass of $1$--$10$~GeV has not been well studied. 
We perform comprehensive studies on the decay of the composite ADM by combining the astrophysical constraints from $e^\pm$ and $\gamma$-ray. 
The constraints from cosmic-ray positron measurements by AMS-02 are the most stringent at $\gtrsim2$~GeV: a lifetime should be larger than the order of $10^{26}$~s, corresponding to the cutoff scale of the portal interaction of about $10^8 \text{--} 10^9 \, \mathrm{GeV}$. 
We also perform the dedicated analysis for the neutrino monoenergetic signals at Super-Kamiokande and Hyper-Kamiokande due to the atmospheric neutrino background in the energy range of our interest. 
\end{abstract}

\maketitle


\section{\label{sec:intro}Introduction\protect}

The existence of dark matter (DM) has been firmly established by astrophysical and cosmological observations. The particle nature of dark matter is still an open question. The DM mass density has been established to be about five times as large as the baryon mass density ($\Omega_\mathrm{DM}\simeq 5 \, \Omega_\mathrm{b}$). The weakly interacting massive particles (WIMPs) have been put forward as the most promising candidate in connection with the origin of electroweak symmetry breaking. Up to today, direct and indirect detection bounds have greatly constrained the WIMP hypothesis.
Asymmetric DM (ADM) is an alternative framework to the WIMP framework, where the DM abundance today is provided by the particle-antiparticle asymmetry of DM particles as with the baryon asymmetry of the Universe (see Refs.~\cite{Nussinov:1985xr, Barr:1990ca, Barr:1991qn, Kaplan:1991ah, Dodelson:1991iv, Kuzmin:1996he, Fujii:2002aj, Kitano:2004sv, Farrar:2005zd, Gudnason:2006ug, Kitano:2008tk, Kaplan:2009ag} for early work, and see also Refs.~\cite{Davoudiasl:2012uw, Petraki:2013wwa, Zurek:2013wia} for reviews).  
Once the number asymmetries of DM and baryons are somehow related to each other during thermal history, the ADM mass is predicted to be in the range of $1$--$10$~GeV.

\begin{figure}
    \centering
    \includegraphics[width=0.45\textwidth]{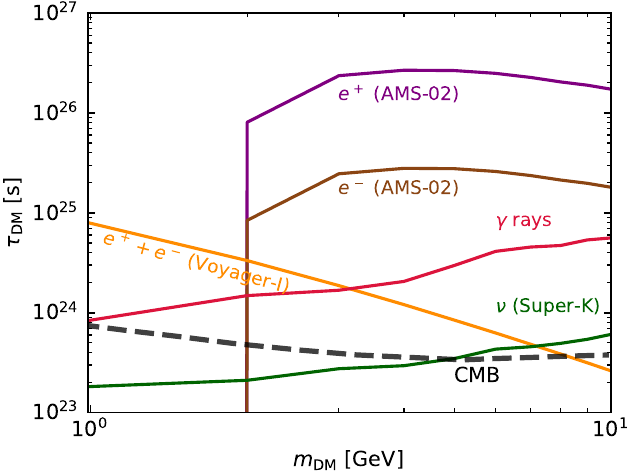}
    \caption{Our summary plot of multimessenger constraints on the lifetime of the composite ADM, assuming the two-body decay ($2 m_{A'} / m_{\pi'}=0.4$) of the dark meson and the vector-like model. Also overlaid are the cosmic microwave background (CMB) constraints obtained in Ref.~\cite{Slatyer:2016qyl}, where a factor of $0.1$ is multiplied since only $\sim10$\% of DM energy goes into $e^+/e^-$. The shaded region shows the region disfavored by each multimessenger component. See text for details.}
    \label{fig:2b_v_mm}
\end{figure}

In this work, we consider composite ADM models with a dark photon portal based on dark quantum chromodynamics (QCD) and dark quantum electrodynamics (QED)~\cite{Ibe:2018juk}.
The asymmetry in the SM sector, generated in the early Universe (via such as the thermal leptogenesis~\cite{Fukugita:1986hr}), is shared to the dark sector through a portal interaction between two sectors. 
The portal interaction plays an important role in not only sharing generated asymmetry in two sectors but also exploring the signature of the composite ADM models.
Since the portal interaction connects ``dark matter number'' with lepton number, it also causes the DM decay into the standard model (SM) leptons: in particular, the dark nucleon decays into the dark meson and antineutrino~\cite{Fukuda:2014xqa}.

Compared to the decaying DM above $10$\,GeV, the decaying DM in the range of $1$--$10$\,GeV, predicted in the composite ADM scenario, has not been studied well.
The ADM decay gives a monochromatic neutrino signal with an energy of roughly half of the DM mass, and the DM lifetime can be constrained by the measurements of neutrino fluxes by detectors such as Super-Kamiokande (Super-K) and Hyper-Kamiokande (Hyper-K). 
The neutrino-line constraint on the decaying DM above $10$\,GeV is discussed in Ref.~\cite{Covi:2009xn}, but one cannot obtain the constraint on the ADM decay by a naive extrapolation of the existing constraints due to the huge atmospheric neutrino background for the mass range of $1$--$10$\,GeV.
Thus, we perform dedicated analyses to derive the neutrino constraints, in particular from the ADM decay with the DM mass of $1$--$10$ GeV, using the expected event rate in Super-K detector \cite{Super-Kamiokande:2002weg} and also project the constraints expected for 10 years of observation by Hyper-K \cite{Abe:2011ts}.
To set a conservative bound on the DM lifetime, we utilize an analogous approach for constraining the neutrino flux with the energy below GeV in  Refs.~\citep{Super-Kamiokande:2002hei, Beacom:2006tt, Yuksel:2007ac, Palomares-Ruiz:2007trf}. 

The dark mesons, the decay products of dark nucleons, can also decay into the SM particles in the presence of dark photons.
The DM lifetime can also be constrained by the measurements of electron/positron fluxes by the AMS-02 detector \cite{AMS:2014xys, Graziani:2017fol} and by using the Voyager data \citep{Stone:2013zlg, Cummings:2016pdr}, and by the isotropic diffuse $\gamma$-ray background measured by Fermi-LAT \cite{Fermi-LAT:2014ryh}, along with the $\gamma$-ray and X-ray data measured by EGRET \citep{EGRET:1997qcq, Strong:2004de}, COMPTEL \citep{Weidenspointner_2000}, and SMM \citep{Watanabe_2000}.
The DM decay directly into an electron-positron pair has been discussed in Ref.~\cite{Boudaud:2016mos}, and the final-state radiation from the DM decay has also been discussed in Ref.~\cite{Essig:2013goa}.
However, it is not straightforward to obtain the constraints on the ADM lifetime from these studies.
Unlike the decaying DM directly into the electron-positron pair \cite{Boudaud:2016mos}, the primary spectrum of $e^-$ and $e^+$ is softened in the composite ADM scenario due to the cascade decay.
The constraints would be sensitive to the change of the primary spectrum because of the energy gap in the existing $e^+/e^-$ data.
This is also similar for the constraints compiling the $\gamma$-ray and the X-ray data.
Besides, the primary spectrum of photons has been obtained by assuming that the inverse Compton scattering is neglected in a bound obtained in Ref.~\cite{Essig:2013goa}.
Thus, we need to compute the primary energy spectrum of the electromagnetic flux from the cascade decay and analyze the multimessenger constraints on the decaying DM in the range of $1$--$10$\,GeV.
Figure~\ref{fig:2b_v_mm} summarizes our multimessenger constraints on the lifetime of composite ADM in a minimal vector-like model and with a two-body decay of the dark meson.  

The outline of this paper is as follows.
In \cref{sec:cADM}, we describe composite ADM models, in particular vector-like realization and chiral realization of the composite ADM.
There, we discuss why the composite ADM mass is in the range of $1$--$10$ GeV, and why we need to introduce the dark photon. 
\cref{sec:decay} provides analytical computation method of the energy spectrum and the primary energy spectrum of the electromagnetic flux from the cascade decay of dark baryons.
We consider not only the cascade two-body decay but also the cascade decay with the three-body decay of the dark pion, where the soft component of the spectrum would increase, and revealed the impacts of the cascade decay that has not been studied in detail.
In \cref{sec:constraints}, we discuss the analysis for each cosmic-ray constraints dedicated to the decaying ADM in the range of $1$--$10$\,GeV and derive multimessenger constraints on ADM decay for various models considered in this work.
\cref{sec:conclusion} is devoted to concluding this study.
We also give detailed discussions in three appendices. 

\section{\label{sec:cADM}Composite Asymmetric Dark Matter}

Compositeness naturally provides key ingredients for the ADM framework when the dark sector where DM resides possesses a confining gauge dynamics as with the QCD in the SM sector~\cite{Gudnason:2006yj, Dietrich:2006cm, Khlopov:2007ic, Khlopov:2008ty, Foadi:2008qv, Mardon:2009gw, Kribs:2009fy, Barbieri:2010mn, Blennow:2010qp, Lewis:2011zb, Appelquist:2013ms, Hietanen:2013fya, Cline:2013zca, Appelquist:2014jch, Hietanen:2014xca, Krnjaic:2014xza, Detmold:2014qqa, Detmold:2014kba, Asano:2014wra, Brod:2014loa, Antipin:2014qva, Hardy:2014mqa, Appelquist:2015yfa, Appelquist:2015zfa, Antipin:2015xia, Hardy:2015boa, Co:2016akw, Dienes:2016vei, Ishida:2016fbp, Lonsdale:2017mzg, Berryman:2017twh,  Gresham:2017zqi, Gresham:2017cvl, Mitridate:2017oky, Gresham:2018anj, Ibe:2018juk, Braaten:2018xuw, Francis:2018xjd, Bai:2018dxf, Redi:2018muu, Chu:2018faw, Mahbubani:2019pij,Hall:2019rld, Tsai:2020vpi, Asadi:2021yml, Asadi:2021pwo, Zhang:2021orr, Bottaro:2021aal, Ibe:2021gil, Hall:2021zsk, Asadi:2022vkc}.
The baryonic particles in the dark sector (dark baryons) are the ADM, and the DM stability is ensured by an accidental number conservation, which is a dark sector counterpart of the baryon number conservation. 
The symmetric component of DM abundance is strongly depleted by the annihilation into the dark-sector mesons (dark meson). 
The GeV mass of DM arises from the dimensional transmutation of the strong dynamics in the dark sector, as with the SM baryons.
After the strong depletion of the symmetric components of the dark baryons, the dark sector contains the large entropy density that is carried by dark mesons.
This can be problematic in cosmology since two sectors are in thermal contact in the early Universe.
Dark photons play an important role in releasing the entropy density: dark mesons are depleted by their annihilation and decay into dark photons. The dark photon decays into pairs of the SM fermions through kinetic mixing with the SM photon, and it releases the entropy of the dark sector into the SM sector.

The DM mass of $1$--$10$\,GeV has not been well explored not only in the indirect detection but also in the direct detection experiments (in particular, using liquid xenon detectors~\citep{XENON:2020kmp, XENON:2022ltv, XENON:2023cxc, LZ:2022lsv, PandaX-4T:2021bab}) due to the experimental threshold for recoil energy.
SM protons can interact with dark baryons through the dark photon portal in our model~\cite{Ibe:2018juk}.%
\footnote{
	The $U(1)_D$ neutral dark baryon can couple to nucleons through the $U(1)_D$ violating mixing between the neutral and charged dark baryons~\cite{Kamada:2021cow} and through the magnetic moment under $U(1)_D$~\cite{Kamada:2020buc}.
}
Due to the DM mass of our interest in the range of $1$--$10$~GeV, it requires the dedicated searches~\cite{CRESST:2019jnq,SuperCDMS:2018gro,DarkSide-50:2022qzh,GlobalArgonDarkMatter:2022ppc,XENON:2022ltv,DarkSide-50:2022qzh,XENON:2019zpr,DarkSide:2022dhx}.

\subsection{A Vector-like Model}

First, we discuss the generic features of the composite ADM with a dark photon portal by considering a model having the $SU(3)_D \times U(1)_D$ gauge dynamics with the vector-like two-flavor dark quarks, which is proposed by Ref.~\cite{Ibe:2018juk}.
We give the particle contents and their charges under the gauge dynamics in \cref{tab:Charge_VL}.
\begin{table}[b]
\caption{\label{tab:Charge_VL}%
Charge assignment in a composite ADM model with two-flavor Weyl dark quarks and the dark Higgs. $SU(3)_D$ and $U(1)_D$ are gauge symmetries of the dark sector, while $U(1)_{B-L}$ is the global symmetry shared with the visible sector.
}
\begin{ruledtabular}
\begin{tabular}{cccc}
		& $SU(3)_D$ & $U(1)_D$ & $U(1)_{B-L}$ \\ \colrule
		$U'$ & $\mathbf{3}$ & $2/3$ & $1/3$ \\
		$\overline U'$ & $\overline{\mathbf{3}}$ & $-2/3$ & $- 1/3$ \\
		$D'$ & $\mathbf{3}$ & $-1/3$ & $1/3$ \\
		$\overline D'$ & $\overline{\mathbf{3}}$ & $1/3$ & $- 1/3$ \\ \colrule
		$\phi_D$ & $\mathbf{1}$ & $1$ & $0$ \\
\end{tabular}
\end{ruledtabular}
\end{table}
The dark quarks (denoted by $U'\,,  \overline U'\,,  D'$\,, and $\overline D'$) are in (anti-)fundamental representations of $SU(3)_D$, have $U(1)_D$ charges similar to the electromagnetic charges of the SM quarks, and carry the same $B-L$ charges as the SM quarks.

The dark quarks are confined into dark hadrons below the dynamical scale $\Lambda_{\mathrm{QCD}'}$ of $SU(3)_D$. 
The Lagrangian possesses the approximate chiral symmetry $SU(2)_L \times SU(2)_R$ as far as the current quark mass and the $U(1)_D$ correction are negligible compared to the dynamical scale. 
Below the dynamical scale, the chiral symmetry is broken to $SU(2)_V$ by chiral condensate,
\begin{equation}
	\langle U' \overline U' + U'^\dag \overline U'^\dag \rangle 
	= \langle D' \overline D' + D'^\dag \overline D'^\dag \rangle 
	= \mathcal{O}(\Lambda_{\mathrm{QCD'}}^3)\,,
\end{equation}
and the low-energy spectrum is written in terms of dark mesons and dark baryons.
The lightest dark mesons are in the triplet representation of $SU(2)_V$, and the lightest dark baryons are in the fundamental representation of $SU(2)_V$.
We adopt a similar name as the SM mesons and the SM baryons for the dark mesons and the dark baryons.
\begin{equation}
	\Pi' = \frac{1}{2}
	\begin{pmatrix}
		\pi'^0  & \sqrt2 \pi'^+  \\
		\sqrt2 \pi'^- & - \pi'^0
	\end{pmatrix} \,, \qquad 
	B' 
	= 
	\begin{pmatrix}
		p' \\
		n' 
	\end{pmatrix} \,.
	\label{eq:meson-baryon_ADM}
\end{equation}
Here, the superscript indicates the $U(1)_D$ charges of the dark hadrons.
The lightest dark baryon is the DM, and its stability is ensured by the $B-L$ number conservation.

\begin{figure}
    \centering
    \includegraphics[width=0.36\textwidth]{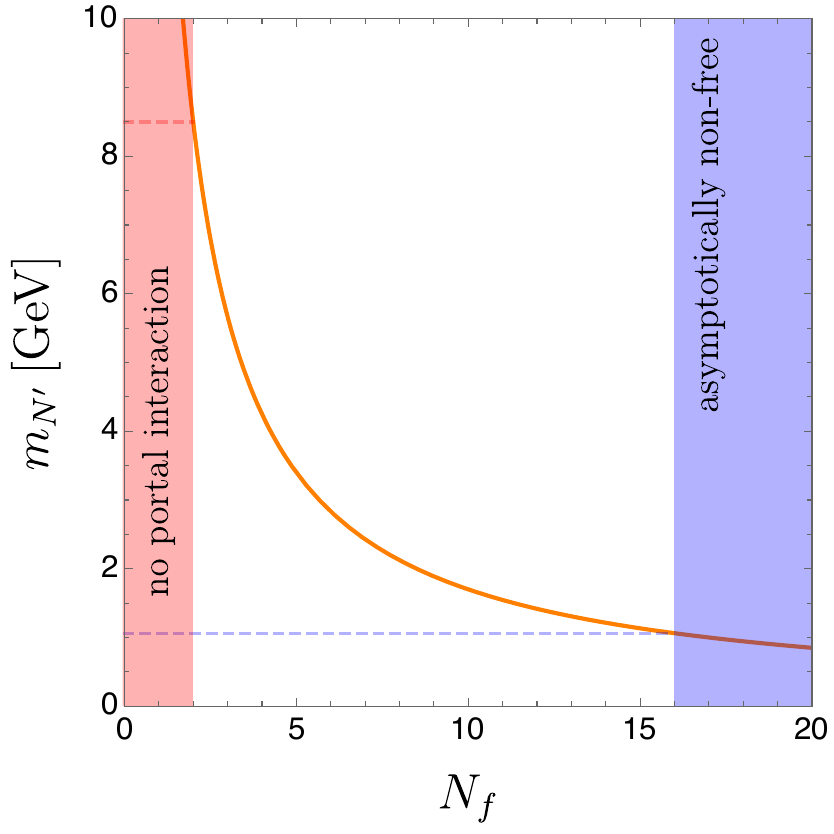}%
    \caption{
	DM mass as a function of the numbers of dark flavors $N_f$ in the vector-like models of the composite ADM. 
	The dark matter mass is predicted to be in the range of 1--10~GeV when its abundance is completely determined by sharing the asymmetries.
	\label{fig:DMmass}
	}
\end{figure}

The baryon asymmetry in the visible sector is assumed to be generated via the thermal leptogenesis~\cite{Fukugita:1986hr} (see Refs.~\cite{Giudice:2003jh,Buchmuller:2005eh,Davidson:2008bu} for reviews).
The generated $B-L$ is shared between the dark sector and the visible sector through portal operators,%
\footnote{
	We assume that the portal interaction preserves the $B-L$ symmetry. 
	Meanwhile, once we consider the ultraviolet realization for the model, this interaction would be generated via integrating the right-handed neutrinos.
	For such a case, we have operators with the $B-L$ charge $-2$, i.e., $\overline U' \overline D' \overline D'$ and $U'^\dag D'^\dag \overline D'$ instead of $U' D' D'$ and $\overline U'^\dag \overline D'^\dag D'$. 
} 
\begin{align}
	\mathcal{L}_\mathrm{portal} \supset & \frac{1}{M^3} (U' D' D') (LH) \nonumber \\
	& + \frac{1}{{M'}^3} (\overline U'^\dag \overline D'^\dag D') (LH) + \mathrm{h.c.}
	\label{eq:Intermediate_Portal}
\end{align}
where $M$ and $M'$ collectively denote mass-dimension one coefficients.
$L$ and $H$ denote the SM lepton doublet and the SM Higgs doublet, respectively.
As shown in Ref.~\cite{Fukuda:2014xqa}, if the asymmetry is fully shared between two sectors, the DM mass is to be $17.0/N_f\,\mathrm{GeV}$. 
$N_f$ denotes the number of dark quark flavors at the temperature where the asymmetries are shared by the portal interactions.
The shared asymmetries are separately conserved in both sectors after the decoupling of the portal interactions. 
Some flavors may be decoupled from the low-energy spectrum if they have Dirac masses. 
We consider minimal numbers of flavors at the low-energy spectrum even though we treat the DM mass as one of the free parameters. 
The DM mass of our interest is in the range of 1 -- 10 GeV as shown in \cref{fig:DMmass}: the case where the DM abundance is completely determined by sharing the asymmetries. 
The one-loop beta function of the dark QCD coupling is proportional to $- (11-2N_f/3)$, and thus the dark QCD is asymptotically free as far as $N_f \leq 16$.
We note that we assume here $N_f$ does not change during the renormalization group running. 
Some of dark quarks at high energy may be decoupled from the low energy spectrum.
The DM mass can be lower than 1 GeV for such a case. 
Meanwhile, to construct the dark-neutral operators, we need to introduce $N_f = 2$ dark quarks $U'$ and $D'$ in the vector-like models at least.

The portal operators also provide the decay of dark baryons below the dark dynamical scale as with the nucleon decay predicted in the grand unified theory: the dark baryon decays into the dark meson and antineutrino~\cite{Fukuda:2014xqa}. 
The portal operators are suppressed by the ultraviolet (UV) scale $M^3$ and $M'^3$. The DM lifetime can be longer than the age of the Universe when the UV scale is larger than $10^8\,\mathrm{GeV}$.
The simplified formula for the decay rate of DM into dark meson and antineutrino is 
\begin{align}
	\Gamma_\mathrm{DM} = \frac{1}{64 \pi} \frac{v^2 m_{N'}}{M_\ast^6} |W|^2 \,,
\end{align}
where $v$ denotes the vacuum expectation value of the SM Higgs and $m_{N'}$ denotes the mass of dark baryons. 
$M_\ast$ includes the information on the cutoff energy scale of the portal operators, say $M$ and $M'$ in \cref{eq:Intermediate_Portal}. 
$W$ describes the matrix element for the dark baryon decay into the dark meson. 
We na\"ively expect the scaling of the matrix element as $|W| \simeq 0.1 \, \mathrm{GeV}^2 (m_{N'}/m_N)^2$ from the lattice results on the nucleon hadron matrix element~\cite{Aoki:2013yxa}, and then the scaling of DM lifetime is given by
\begin{align}
	\tau_\mathrm{DM} 
	\simeq 10^{26} \, \mathrm{s} \left(\frac{M_\ast}{2 \times 10^9 \, \mathrm{GeV}}\right)^6 \left(\frac{10\,\mathrm{GeV}}{m_{N'}}\right)^5 \,.
\end{align}

\subsection{A Chiral Model}

In order to break $U(1)_D$, we introduce the dark Higgs boson in the original model~\cite{Ibe:2018juk}, or we may consider the St\"uckelberg mechanism~\cite{Stueckelberg:1938hvi,Ruegg:2003ps,Kors:2004dx,Feldman:2007wj}.
We have to require tuning of parameters to obtain the dark photon mass of $O(10^1 \text{--} 10^2) \, \mathrm{MeV}$ even though the $U(1)_D$ breaking scale has nothing to do with the chiral symmetry breaking.
Another possibility is the dynamical generation of the dark photon mass via the chiral symmetry breaking~\cite{Harigaya:2016rwr,Co:2016akw}. 

We discuss a chiral realization for the composite ADM~\cite{Ibe:2021gil}. 
We assign different $U(1)_D$ charges to a Weyl fermion and its chiral partner (see \cref{tab:Charge_Ch} for the charge assignment). 
\begin{table}[b]
	\caption{
	Charge assignment in a chiral composite ADM model with three-flavor Weyl dark quarks.
	The gauge and global symmetries are the same as \cref{tab:Charge_VL}, but the $U(1)_D$ charge assignment differs.
	\label{tab:Charge_Ch}
	}
    \begin{ruledtabular}
	\begin{tabular}{cccc}
		& $SU(3)_D$ & $U(1)_D$ & $U(1)_{B-L}$ \\ \colrule
		$U'$ & $\mathbf{3}$ & $1$ & $1/3$ \\
		$\overline U'$ & $\overline{\mathbf{3}}$ & $-a$ & $- 1/3$ \\
		$D'$ & $\mathbf{3}$ & $-1$ & $1/3$ \\
		$\overline D'$ & $\overline{\mathbf{3}}$ & $a$ & $- 1/3$ \\
        $S'$ & $\mathbf{3}$ & $0$ & $1/3$ \\
		$\overline S'$ & $\overline{\mathbf{3}}$ & $0$ & $- 1/3$ \\ 
	\end{tabular}
    \end{ruledtabular}
\end{table}
We also introduce new vector-like dark quarks $S'$ and $\overline S'$ in order to have an appropriate portal interaction.
The theory is chiral as $a \neq 1$, and hence, the Dirac mass terms for dark quarks are forbidden.
Below the dynamical scale $\Lambda_{\mathrm{QCD'}}$ of $SU(3)_D$, the dark quark condensation given by 
\begin{align}
	\langle U' \overline U' + U'^\dag \overline U'^\dag \rangle 
	& = \langle D' \overline D' + D'^\dag \overline D'^\dag \rangle \nonumber \\
	& = \langle S' \overline S' + S'^\dag \overline S'^\dag \rangle 
	= \mathcal{O}(\Lambda_{\mathrm{QCD'}}^3)\,,
\end{align}
breaks the chiral symmetry,%
\footnote{
We assume a similar chiral condensate for $S' \overline S'$ even though a net $U(1)_D$ charge for $S' \overline S'$ differs from that for the other quark bilinear.
} 
and the $U(1)_D$ is spontaneously broken by this condensation. 
Therefore, one of the dark neutral mesons is absorbed as the longitudinal degrees of freedom of the massive dark photons.
As in \cref{eq:meson-baryon_ADM}, we adopt a similar name as the SM hadrons for the dark hadrons.
\begin{align}
	\Pi'
	& = \frac{1}{2}
	\begin{pmatrix}
		\frac{1}{\sqrt3} \eta' & \sqrt2 \pi' & \sqrt2 K'_1 \\
		\sqrt2 \pi'^\dag & \frac{1}{\sqrt3} \eta' & \sqrt{2} K'_2 \\
		\sqrt{2} K'^\dag_1 & \sqrt{2} K'^\dag_2 & -\frac{2}{\sqrt3}\eta'
	\end{pmatrix} \,, \\
	B' 
	& = \frac{1}{2}
	\begin{pmatrix}
		\Sigma'_3 + \frac{1}{\sqrt3} \Lambda'  & \sqrt{2} \Sigma' & \sqrt{2} p' \\
		\sqrt{2} \Sigma'^\dag & -\Sigma'_3 + \frac{1}{\sqrt3} \Lambda' & \sqrt{2} n' \\
		\sqrt{2} \Xi'_2 & \sqrt{2} \Xi'_1 & - \frac{2}{\sqrt3} \Lambda'
	\end{pmatrix} \,.
	\label{eq:meson-baryon_CADM}
\end{align}
We note that the $U(1)_D$ charges of the dark hadrons are not same as the electromagnetic charges of the corresponding SM hadrons. 

The portal interaction sharing the generated $B-L$ asymmetry is similarly given by the higher-dimensional operators. 
The following operators are allowed due to our $U(1)_D$ charge assignments: 
\begin{align}
	\mathcal{L}_\mathrm{portal} \supset & \frac{1}{M^3_\mathrm{DS}} (U' D' S') (LH) \nonumber \\
	& + \frac{1}{{M'}_\mathrm{DS}^3} (\overline U'^\dag \overline D'^\dag S') (LH)
	+ \mathrm{h.c.} \,, 
	\label{eq:Intermediate_Portal_Ch}
\end{align}
where $M_\mathrm{DS}$ and $M'_\mathrm{DS}$ denote mass-dimension one coefficients.
As with the previous model, the dark baryons decay into dark mesons and antineutrinos via the portal interactions.%
\footnote{
	One can decouple the vector-like dark quarks $S'$ and $\overline S'$ from the low energy spectrum to realize a chiral nature of the model even at low energy.
	For such a case, however, dark baryons are stable.
} 
Since we need to have three flavors to construct the portal interaction, the DM mass is below 5.7 GeV if the shared asymmetry determines the total DM abundance. 
Since one of the $U(1)_D$-neutral dark mesons is absorbed as the dark photons, we also have the decay channel that directly emits a longitudinal mode of the dark photon in the chiral model.

\section{\label{sec:decay}Decay of Dark Baryons}

The dark baryons decay into dark mesons (or a longitudinal component of the dark photon) and the SM antineutrinos through the portal interactions~\cref{eq:Intermediate_Portal,eq:Intermediate_Portal_Ch} below the dark confining scale, as we mentioned in the previous section.
We assume that the dark-baryon decay into spin-one dark mesons (such as dark-sector counterpart of $\rho$ mesons) or more than two mesons is negligible.%
\footnote{
	In the context of the grand unification theory, nucleons can decay into three-body final states (two pions and an antilepton). 
	Due to the final state interaction between pions, the three-body decay rate can be comparable to the two-body decay rate~\cite{Wise:1980ch}.
	We ignore these channels in this study, but it is worthwhile to investigate other decay channels of dark baryons. 
}
The final-state electromagnetic particles may also be produced from the multi-step cascade decay of the dark baryons in the presence of the kinetic mixing between the dark photon and the SM photon.
Indeed, the $U(1)_D$-neutral dark baryons decay into neutral dark mesons and the SM antineutrinos.
The neutral dark mesons promptly decay into two dark photons via the chiral anomaly in the dark sector. 
The dark photon finally decays into the electromagnetic particles through the kinetic mixing. 
The dark photon lifetime depends on the kinetic mixing, and we assume the lifetime is smaller than $\mathcal{O}(1)\,s$ (corresponding to the kinetic mixing $\epsilon \gtrsim 10^{-10}$ for the dark photon mass of $100\,\mathrm{MeV}$) in order to avoid the late-time reheating of the electromagnetic plasma.

Before discussing the electromagnetic fluxes from the cascade decay of dark baryons, we comment on numerical coefficients specific to the number of dark flavors. 
All dark baryons do not leave electromagnetic signals from the cascade decay. 
A portion of dark baryons in the Universe can decay into dark neutral mesons decaying into dark photons through chiral anomaly. 
On the other hand, the remaining portion of dark baryons can decay into darkly-charged mesons and antineutrinos. 
The darkly-charged mesons are stable,%
\footnote{
	Some of the darkly-charged mesons may decay into two dark photons through the $U(1)_D$ breaking vacuum expectation value. 
	The decay rate of the darkly-charged mesons into the electromagnetic channel is suppressed by the mixing between the dark-neutral and darkly-charged mesons.
	We expect that the branching fraction from the darkly-charged mesons is very tiny, and we ignore these channels in this study. 
}%
and this decay mode leaves missing particle and antineutrino signals. 
We should multiply the energy spectra of the electromagnetic particles from a single dark baryon, which will be derived in this section, by the factor of the decay rate with the dark neutral mesons in the final state to the sum of all decay rates of dark baryons.
In \cref{app:DR}, we discuss the decay rate of dark baryons in the vector-like composite ADM and in the chiral composite ADM. 

\subsection{Two-body decay of the dark meson}

We consider the electromagnetic flux from the decay of a dark baryon. 
When the decay of a dark neutral meson into two dark photons is kinematically allowed (namely, $m_{\pi'} > 2 m_{A'}$), the dark baryons decay into the electromagnetic particles through the multi-step two-body decay of dark meson. 
The dark neutron decays into the SM particles through $n' \to \pi'^0 +\overline \nu$ and $ \pi'^0 \to 2 A' \to 2 (e^- e^+ (\gamma))$ in the vector-like composite ADM, while $\Sigma_3'$ and $\Lambda'$ decay into $\eta'$ and $\eta'$ decays into the electromagnetic channel via $\eta' \to 2 A' \to 2 (e^- e^+ (\gamma))$ in the chiral composite ADM.

The primary energy spectra of an electron (positron) and a photon from the dark photon $A'$ decay are
\begin{align}
	& \frac{d N_{e^-(e^+)}}{dx_0} = \delta(1-x_0) \,, \nonumber \\
	& \frac{d N_\gamma}{dx_0} 
	= 2 \times \frac{\alpha}{2\pi} \frac{1+(1-x_0)^2}{x_0} \left(\ln \frac{4(1-x_0)}{\epsilon_0^2} - 1\right) \,, 
	\label{eq:zeroth_spectrum}
\end{align}
where $x_0 = 2E_0/m_{A'}$ and $\epsilon_0 = 2 m_e/m_{A'}$ with the energy $E_0$ of the electron (positron) or the photon in the $A'$ rest frame.
We use the Altarelli-Parisi formula for the photon spectrum from the final-state radiation~\cite{Mardon:2009rc}, and $\alpha$ denotes the electromagnetic fine structure constant.
$N_{e^-(e^+)}$ and $N_{\gamma}$ give the (average) number of electrons and photons per $A'$ decay, respectively. 
In particular, we adopt the normalization that the integral of $d N_{e^-(e^+)}/dx_0$ over $x_0 \in [0, 1]$ gives one. 

Next, the spectrum is boosted to the rest frame of dark baryons. 
We give a detailed discussion of the energy spectra from the cascade decay in \cref{app:cascade}.
When the $\pi'$ decay into two dark photons is kinematically allowed, the following formula gives the spectrum in the $\pi'$ rest frame.
\begin{align}
	& \frac{d N_\psi}{dx_1} =
	2 \int_{-1}^1 d \cos\theta_{0} \int_{\epsilon_0}^1dx_{0} \frac{d N_\psi}{dx_{0}} f_{0} (\cos\theta_{0}) \nonumber \\ 
	& \quad \times \delta\left[x_1 - \frac{1}{2} \left(x_{0} + \sqrt{(1-\epsilon_1^2)(x_{0}^2 - \tilde \epsilon_1^2)} \cos\theta_{0}\right) \right] \,.
	\label{eq:spectrum_pirest}
\end{align}
Here, $x_1 = 2 E_1/m_{\pi'}$, $\epsilon_1 = 2 m_{A'}/m_{\pi'}$, and $\tilde \epsilon_1 = 2 m_{\psi}/m_{\pi'}$ with $E_1$ the energy of a particle $\psi = e^\pm \,, \gamma$ in the $\pi'$ rest frame.
A prefactor two implies the multiplicity of $A'$ in the $\pi'$ decay.
The function $f_0$ denotes the angular distribution of the $A'$ decay.
For the transverse mode of the dark photon, we take 
\begin{equation}
  f_0(\cos \theta) = \frac{3}{8} (1 + \cos^2 \theta) \,.
\end{equation}
The delta function gives a relation of the energies in the rest frames of $\pi'$ and $A'$.
Then, we obtain the primary spectrum in the rest frame of the dark baryons by boosting it, again.
The spectrum in the $N'$ rest frame is given by
\begin{align}
	& \frac{d N_\psi}{dx_2} 
	= \int^1_{-1} d \cos\theta_{1} \int^1_{\tilde\epsilon_1} dx_{1} \frac{d N_\psi}{dx_1} \times \frac12 \nonumber \\
	& \quad \times \delta\left(x_2 - \frac{1+\epsilon_2^2}{2} x_1 - \frac{1-\epsilon_2^2}{2} \sqrt{x_1^2 -\tilde\epsilon_2^2}\cos\theta_{1}\right) \,.
	\label{eq:Spectrum_Nrest}
\end{align}
Here, we assume that antineutrinos in the decay products of the dark baryons are massless. 
We define the dimensionless parameters: $x_2 = 2 E_2/m_{N'}$, $\epsilon_2 = m_{\pi'}/m_{N'}$, and $\tilde \epsilon_2 = 2 m_{\psi}/m_{N'}$ with $E_2$ the energy of $\psi$ in the $N'$ rest frame.
The factor of $1/2$ comes from the angular distribution of the $N'$ decay.
The mass spectrum in the dark sector is usually hierarchical in the composite ADM models; the dark baryons are much heavier than the dark mesons and the dark photon, in other words, $\epsilon_2 \ll 1$ and $\tilde \epsilon_1 \ll 1$. 
In such a case, we may simplify the energy spectrum in \cref{eq:Spectrum_Nrest} as follows.
\begin{equation}
	\frac{d N_\psi}{dx_2} 
	\simeq \int^1_{x_2} \frac{dx_{1}}{x_1} \frac{d N_\psi}{dx_1} \,.
	\label{eq:Spectrum_Nrest_simplified}
\end{equation}

In the chiral model, one of the dark mesons is absorbed as the longitudinal mode of the dark photon. 
Therefore, the dark photon is directly emitted from the dark baryon in this model: $\Sigma_3' (\Lambda') \to A' + \overline \nu$ in the model with $N_f = 3$. 
We use a simplified formula for this mode since the dark photon is sufficiently lighter than the dark baryons.
\begin{equation}
	\frac{d N_\psi}{dx_2} 
	\simeq \int^1_{x_2} \frac{dx_{0}}{x_0} \frac{d N_\psi}{dx_0} 2 \widetilde f_0 \left( \frac{2 x_2}{x_0} - 1\right)\,.
	\label{eq:Spectrum_Nrest_simplified_cADM}
\end{equation}
Here, a relation of the energies in the rest frames of $A'$ and $N'$ determines the argument of the angular distribution $\widetilde f_0$.
We note again that $x_2$ is the energy of the final state particles normalized by $m_{N'}$ in the $N'$ rest frame, while $x_0$ is the energy normalized by $m_{A'}$ in the $A'$ rest frame.
We use the following angular distribution of $A'$ decay for the longitudinal mode of the dark photon.
\begin{equation}
  \widetilde f_0(\cos \theta) = \frac{3}{4} (1 - \cos^2 \theta) \,.
\end{equation}

\subsection{Three-body decay of the dark meson}

\begin{figure*}
    \centering
    \includegraphics[width=0.75\textwidth]{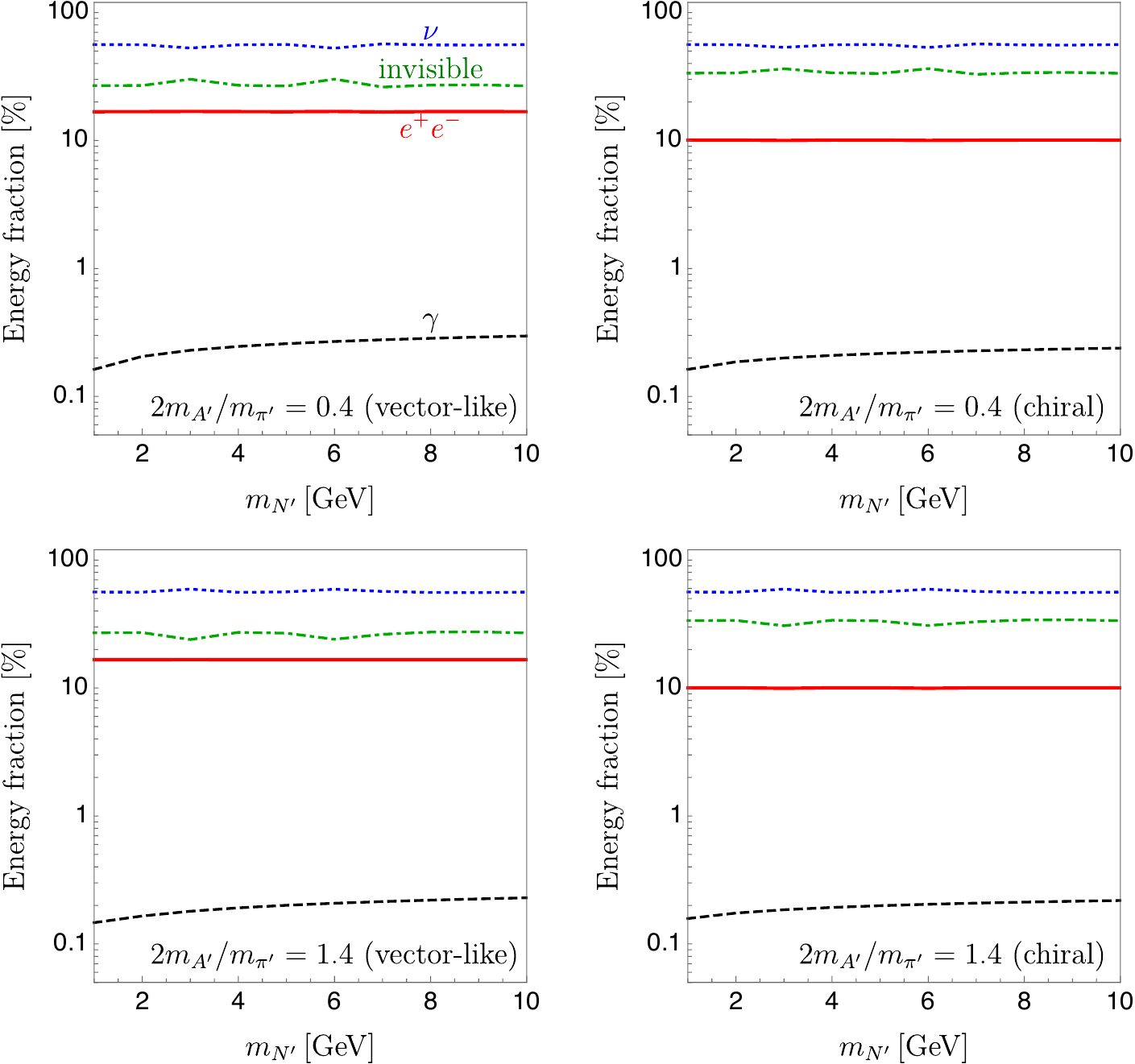}
    \caption{
		Energy fractions of the dark matter decay: 
		electron and positron (red-solid), $\gamma$-ray (black-dashed), all neutrinos (blue-dotted), and invisible channels (green dot-dashed). 
		The two-body decay of a dark meson $\pi' \to A' A'$ is kinematically allowed in the top panels, while the three-body decay of a dark meson $\pi' \to A' e^+ e^- (\gamma)$ is only allowed in the bottom panels. 
 	}
    \label{fig:fraction}
\end{figure*}

We have discussed the case where the $\pi'$ decay into two dark photons is kinematically allowed, namely $m_{\pi'} > 2 m_{A'}$. 
Meanwhile, when the mass of the dark photon is close to that of the dark mesons, $\pi'$ dominantly decays into $A'$ and a pair of $e^+ e^- (\gamma)$ via an off-shell dark photon.
Both vector-like and chiral models have the decay: $n' \to \pi'^0 +\overline \nu$ and $ \pi'^0 \to A' e^- e^+ (\gamma) \to 2 (e^- e^+ (\gamma))$ in the vector-like composite ADM, while $\Sigma_3'$ and $\Lambda'$ decay into $\eta'$ and $\eta'$ decays into the electromagnetic channel via $\eta' \to A' e^- e^+ (\gamma) \to 2 (e^- e^+ (\gamma))$ in the chiral composite ADM.
In \cref{app:cascade}, we give detailed discussions on the cascade decay with an off-shell dark photon.

Now, we consider the decay $\pi' \to A' + e^- + e^+$ in the $\pi'$ rest frame. The energy spectrum of the decay is described by two energy quantities normalized by the $\pi'$ mass: the energy of $A'$, $\xi_1 = 2 E_{A'}/m_{\pi'}$, and that of the electron, $\xi_2 = 2 E_{e}/m_{\pi'}$ in the $\pi'$ rest frame.
We find the energy spectrum of the three-body decay by ignoring the electron mass as follows. 
\begin{align}
	& \frac{1}{\Gamma_{A'e^+e^-}} \frac{d \Gamma_{A'e^+e^-}}{d \xi_1 d \xi_2} = 
	\frac{1}{f(\overline \epsilon_1)}
	\frac{1 + \overline \epsilon_1^2 - \xi_1}{(1-\xi_1)^2} \nonumber \\
	& \times \left[ \xi_1^2 + 2 \xi_1 (\xi_2 -1) + 2 (\xi_2^2 - 2 \xi_2 - \overline \epsilon_1^2 + 1) \right]\,.
	\label{eq:spectrum_3body}
\end{align}
Here, $\overline \epsilon_1 = m_{A'}/m_{\pi'}$. 
The kinematics of the final state particles determines the range of parameters $\xi_1$ and $\xi_2$. 
The function $f$ is given by 
\begin{align}
	& f(t) = - \frac{1}{9} (t^6 + 63 t^4 - 81 t^2 - 17) \nonumber \\
	& \quad + \frac{2}{3} (6 t^4 - 9 t^2 + 1) \ln t \nonumber \\ 
	& \quad + \frac{2(28 t^4 - 11 t^2 + 1)}{3 \sqrt{4 t^2 - 1}} \tan^{-1} \frac{3 t^2 - 1}{(t^2 - 1)\sqrt{4 t^2 - 1}} \,.
	  \label{eq:fx_3body}
  \end{align}

The electron/photon spectrum consists of two contributions: one is from the final state electrons/photon, and another is from the decay of the final state $A'$.
The former contribution is easily obtained by integrating the energy spectrum over the kinematic variable $\xi_1$ and by boosting it to the $N'$ rest frame. 
As for the electron spectrum, we have 
\begin{align}
  \frac{d N_{e^-(e^+)}}{d x_2} & = \int^{1-\overline \epsilon_1^2}_{x_2} \frac{d \xi_2}{\xi_2}  \frac{d N_{e^-(e^+)}}{d \xi_2} \,, \\ 
	\frac{d N_{e^-(e^+)}}{d \xi_2} & \equiv \frac{1}{\Gamma_{A'e^+e^-}} \int_{\xi_{1:\mathrm{min}}}^{\xi_{1:\mathrm{max}}} d \xi_1  \frac{d \Gamma_{A'e^+e^-}}{d \xi_1 d \xi_2} \,.
\end{align}
Here, we define $x_2 = 2 E_2/m_{N'}$ ($E_2$ denotes the energy of electron in the $N'$ rest frame), again.
For the fixed energy of electron, $\xi_2$, the integration range of $\xi_1$ is 
\begin{equation}
  \xi_{1:\mathrm{min}} = \frac{1 + \overline \epsilon_1^2 - 2 \xi_2 + \xi_2^2}{(1 - \xi_2)} \,, \quad 
  \xi_{1:\mathrm{max}} = 1 + \overline \epsilon_1^2 \,.
\end{equation}
After the integration on $\xi_1$, $\xi_2$ varies within the range of $[0, 1-\overline \epsilon_1^2]$.
We use the approximation that $N'$ is much heavier than $\pi'$ to get the spectrum in the $N'$ rest frame.
In a similar way, we get the photon spectrum as follows. 
\begin{align}
  \frac{d N_{\gamma}}{d x_2} & = \int^{1}_{x_2} \frac{d x_0}{x_0} \frac{d N_{\gamma}}{d x_0} \,, \\
  \frac{d N_{\gamma}}{d x_0} & = 2 \int d \xi_2 d \xi_1 
  \frac{1}{\Gamma_{A'e^+e^-}} \frac{d \Gamma_{A'e^+e^-}}{d \xi_1 d \xi_2} \frac{d N_{\gamma}}{d r_0} \,, \label{eq:AP_3body_pirest} \\
  \frac{d N_{\gamma}}{d r_0} & = \frac{\alpha}{2 \pi} \frac{1+(1-r_0)^2}{r_0} \nonumber \\ 
  & \qquad \times \left( \ln  \frac{4(1-r_0)(1-\xi_1 + \overline \epsilon_1^2)}{\overline \epsilon_0^2} - 1\right) \,.
  \label{eq:AP_3body}
\end{align}
Here, $r_0 \equiv x_0/\xi_2 (= E_0/E_e)$ denotes the energy fraction carried by photon from the final state electron (or positron), and $\overline \epsilon_0 = 2 m_e/m_{\pi'}$.
One obtains the photon spectrum (\ref{eq:AP_3body_pirest}) in the $\pi'$ rest frame by multiplying the energy spectrum of the decay process to the splitting function and integrating over electron (positron) energy $\xi_2$ and dark photon energy $\xi_1$. 
The factor two indicates the sum of photon spectra from electron and positron. 

For the latter contribution, we first get the energy spectrum of $A'$ by integrating over the kinematic variable $\xi_2$.
\begin{align}
  \frac{d N_{A'}}{d \xi_1} \equiv \frac{1}{\Gamma_{A'e^+e^-}} \int_{\xi_{2:\mathrm{min}}}^{\xi_{2:\mathrm{max}}} d \xi_2  \frac{d \Gamma_{A'e^+e^-}}{d \xi_1 d \xi_2} \,.
\end{align}
Here, the integration range of $\xi_2$ is given by
\begin{equation}
  \xi_{2:\mathrm{max/min}} = \frac12 \left(2 - \xi_1 \pm \sqrt{\xi_1^2 - \overline \epsilon_1^2} \right)\,.
\end{equation}
This spectrum is almost flat in the range of $\xi_1 \in [2 \overline \epsilon_1, 1+\overline \epsilon_1^2]$.
Approaching the boundaries of the range, the spectrum is steeply diminished.
Then, we obtain the energy spectra of the electron-positron pair and photon from the decay of on-shell $A'$ in the $\pi'$ rest frame by the integral
\begin{align}
  \frac{dN_{\psi}}{dx_1} & = \int_{\epsilon_1}^{1+\overline \epsilon_1^2} d \xi_1 \int_{-1}^1 d \cos\theta \int_{\tilde \epsilon_0}^1 d x_0 \nonumber \\
	& \times \frac{dN_{\psi}}{dx_0}(x_0,\epsilon_0)\frac{d N_{A'}}{d \xi_1} (\xi_1,\epsilon_1) f(\cos\theta) \nonumber \\
  & \times \delta \left( x_1 - \frac{\xi_1}{2} x_0 + \frac12 \sqrt{\xi_1^2 - \epsilon_1^2} \sqrt{x_0^2 - \tilde \epsilon_1^2} \cos\theta\right) \,.
\end{align}
Here, $dN_\psi/d x_0$ denotes the primary spectra of an electron and photon from the dark photon given by \cref{eq:zeroth_spectrum}.
When taking $d N_{A'}/d \xi_1 \to \delta (\xi_1 - 1)$, this reproduces the spectrum for two-body cascade decay given by \cref{eq:spectrum_pirest}.
The spectra in the $N'$ rest frame are obtained by boosting again, and its simplified formula is shown in \cref{eq:Spectrum_Nrest_simplified_cADM}.
In summary, as $m_{A'} < m_{\pi'} < 2 m_{A'}$, $\pi'$ decay into two dark photons is not kinematically allowed, and the electron/photon energy spectra are given by the sum of two contributions: 
\begin{equation}
  \frac{dN_{\psi}}{dx_2} = \left. \frac{dN_{\psi}}{dx_2} \right|_{\psi} 
	+ \left. \frac{dN_{\psi}}{dx_2} \right|_{A'} \,,
\end{equation}
where the former denotes the spectra directly from $\pi' \to A' e^+e^-(\gamma)$ and the second term is the contribution from the decay of dark photon. 

At the end of this section, we comment on the energy fractions of $e^\pm$\,, $\gamma$\,, $\nu$\,, and invisible particles (darkly-charged dark mesons) from the ADM decay.
In the following, we consider four benchmark models: mass ratios of $2 m_{A'} / m_{\pi'} = 0.4 \,, 1.4$ for a minimal model with two-flavor vector-like dark quarks and for a minimal model with three-flavor chiral dark quarks.
We assume the universal DM mass $m_{N'}$ and the mass ratio $m_{\pi'}/m_{N'} = 0.1$ for $2 m_{A'} / m_{\pi'} = 0.4$ and $m_{\pi'}/m_{N'} = 0.03$ for $2 m_{A'} / m_{\pi'} = 1.4$. 
With these parameters, the dark photon mass $m_{A'}$ is within the range of $20\,\mathrm{MeV} \text{--} 200 \,\mathrm{MeV}$ for $m_{N'} = 1 \text{--} 10 \, \mathrm{GeV}$.
This dark photon mass is free from the $N_f$ constraint of the dark photon parameters~\cite{Ibe:2018juk,Ibe:2019gpv}. 
Other final states such as $\mu^+ \mu^-$ and $\pi^+ \pi^-$ are not kinematically opened for this mass spectrum. 
When the $A'$ decay into $\mu^+ \mu^-$ and $\pi^+ \pi^-$ is allowed, the final states promptly decay into $e^+$ and $e^-$ with emitting neutrinos, and hence most of the energy is carried out by neutrinos. 
It is expected that the constraints from the astrophysical observations of the $e^+ \,, e^-\,,$ and $\gamma$-ray get weaker once the other final states are kinematically allowed. 

\cref{fig:fraction} shows the DM mass dependence of the energy fractions for each benchmark model. 
The energy fractions hardly depend on the DM mass and do not significantly change among the benchmark models. 
The SM neutrinos (blue-dotted line) take away half of the total energy since the DM decay via the portal interaction always includes neutrinos as the final state. 
Therefore, the neutrino signal from the DM decay is monochromatic with an energy equivalent to half of the DM mass, and it does not significantly change among benchmark models. 
A quarter is taken away by the darkly charged dark mesons (invisible particles: green dot-dashed line).
The rest of the energy is carried by a pair of electrons and positron (red-solid line) and $\gamma$ rays (black-dashed lines), which arise from the decay of the dark-neutral dark mesons.

\section{\label{sec:constraints}Multimessenger constraints \protect}

The decay of composite ADM results in the production of electrons/positrons ($e^\pm$), $\gamma$ rays, and neutrinos ($\nu$) 
up to 10 GeV. Terrestrial and space-based multimessenger detectors operating within this energy range may observe the flux of these particles. In this section, we calculate the expected astrophysical signals resulting from the decay of composite ADM. We utilize existing multimessenger data to put 95\% confidence level (C.L.) constraints on the lifetime of composite ADM. We consider two values for the ratio between the dark meson and dark photon masses, $2m_{A'}/m_{\pi'} =0.4$ and 1.4, corresponding to the two-body and three-body decay of the dark meson, respectively. For each decay mode, minimal models of vector-like and chiral quarks lead to four specific cases examined in this study.

\subsection{\label{subsec:dm_decay}ADM decay spectrum \protect}

We consider a dark matter density distribution in our Galaxy given by a generalized form of the Navarro-Frenk-White (NFW) profile \cite{Navarro:1996gj, Navarro:2003ew},
\begin{equation}
    \rho_{\rm DM}(R) = \rho_{\rm sc} \bigg(\dfrac{R}{R_{\rm sc}}\bigg)^{-\gamma} \bigg(\dfrac{1+R/R_c}{1+R_{\rm sc}/R_c}\bigg)^{-(3-\gamma)},
\end{equation}
where $\gamma = 1.2$ and the core radius $R_c=20$ kpc. We consider the distance of the Sun from the Galactic center to be $R_{\rm sc}=8.5$ kpc, and the DM density in the solar neighborhood to be $\rho_{\rm sc} c^2=0.43$ GeV/cm$^3$ \cite{Salucci:2010qr}. The boundary of the halo in the direction $\theta$ is given as
\begin{equation}
s_{\rm max}(\theta) = R_{\rm sc}\cos\theta + \sqrt{R_h^2 - R_{\rm sc}^2\sin^2\theta}.
\end{equation}
We take $R_h=100$ kpc as the size of the Galactic halo. 

$\gamma$ rays and neutrinos propagate along geodesics in space. The line-of-sight integration of their fluxes originating from the direction $\theta$ is given as
\begin{align}
\Phi(E,\theta) &= \dfrac{1}{4\pi m_{\rm \chi}\tau_{\rm \chi}} \dfrac{dN_s}{dE} \int_{0}^{s_{\rm max}(\theta)}  \rho_\chi(R(s))ds \nonumber \\
&= \dfrac{\rho_{\rm sc} R_{\rm sc}}{4\pi m_\chi\tau_\chi} \dfrac{dN_s}{dE}\mathcal{J}^{\rm dec}(\theta). \label{eqn:los}
\end{align}
The Galactic contribution is obtained by performing the following integration up to $\theta=\pi$:
\begin{align}
\Phi_{\rm G}(E, \leq\theta) = \dfrac{\rho_{\rm sc} R_{\rm sc}}{4\pi m_\chi\tau_\chi} \dfrac{dN_s}{dE} \mathcal{J}^{\rm dec}_{\Omega}\\
\text{where, \ \ \ } \mathcal{J}_{\Omega}^{\rm dec} = \dfrac{2\pi}{\Omega}\int_{0}^{\theta}  \sin\theta d\theta \mathcal{J}^{\rm dec}(\theta). \label{eqn:phi_gal}
\end{align}
Here, $\Omega=2\pi(1-\cos\theta)$ is the solid angle of the field of view, and the integration in \cref{eqn:los} is carried out by changing the variable from line-of-sight coordinate $s$ to Galactocentric distance $R$. 

It suffices to consider only the Galactic component of $\gamma$-rays for the DM mass range we consider in this study since the extragalactic flux is substantially lower in comparison. Whereas, for decaying DM, the extragalactic contribution is significant for neutrinos, and it is known to be comparable to the Galactic contribution~\cite{Covi:2009xn,Murase:2012xs}. We consider a uniform DM density distribution up to redshift $z=20$ for the extragalactic component. In this case, the neutrino flux from the decay of extragalactic ADM is given as
\begin{align}
    \Phi^{\rm dec}_{\rm EG}(E)=
    \dfrac{\Omega_{\rm DM}\rho_c}{4\pi m_{\rm DM} \tau_{\rm DM}}\int dz \bigg|\dfrac{dt}{dz}\bigg| F(z) \dfrac{dN_s^{\rm ob}}{dE}(z_s=z),
    \label{eqn:extragalactic}
\end{align}
where $dN_s^{\rm ob}/dE$ is the all-flavor neutrino spectrum observed on Earth after propagation of the prompt DM decay spectrum originating at redshift $z$. Here, $\rho_c$ is the critical density in a flat Friedmann–Lemaître–Robertson–Walker universe, $|dt/dz|$ is the cosmological line element, and $\rho_{\rm DM} = \Omega_{\rm DM}\rho_c$. We take $\Omega_{\rm DM} h^2=0.113$ and $\rho_c h^{-2}=1.05\times10^{-5}$~GeV~cm$^{-3}$, where $h=0.673$ is the dimensionless Hubble constant \citep{Planck:2018vyg}. The factor $F(z)$ is the redshift evolution of the source population for extragalactic DM, which is assumed to be unity for the decaying ADM.

\subsection{Cosmic-ray propagation}

\begin{figure}[t!]
    \includegraphics[width=0.45\textwidth]{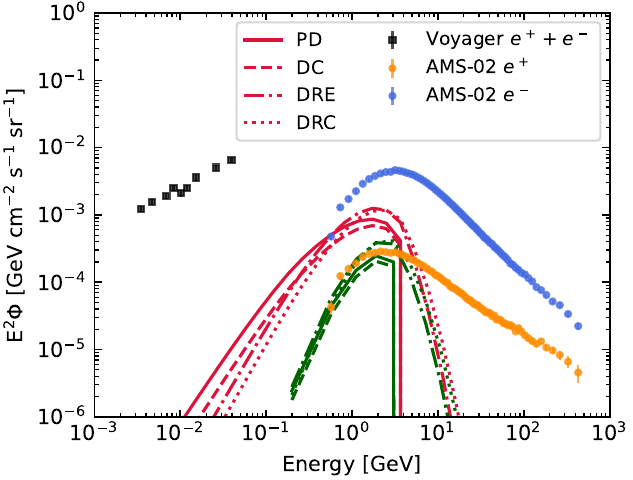}
        \caption{All-sky $e^\pm$ fluxes for the two-body vector-like decay case with $m_{\rm DM}=10$~GeV and $2m_A/m_\pi=0.4$. We use $\tau_\mathrm{DM} = 10^{26}$~s for the demonstration. Four propagation scenarios are shown: PD, DC, DRE, and DRC (see \cref{tab:propagation}). The red lines show the predicted spectra without solar modulation, while the green lines show the predicted spectra at Earth (with modulation).
        The flux measurements from AMS-02 \cite{AMS:2014xys, Graziani:2017fol} and Voyager-1 \citep{Stone:2013zlg, Cummings:2016pdr} are shown.}
    \label{fig:epm_spectrum}
\end{figure}

\begin{figure}[t!]
    \includegraphics[width=0.45\textwidth]{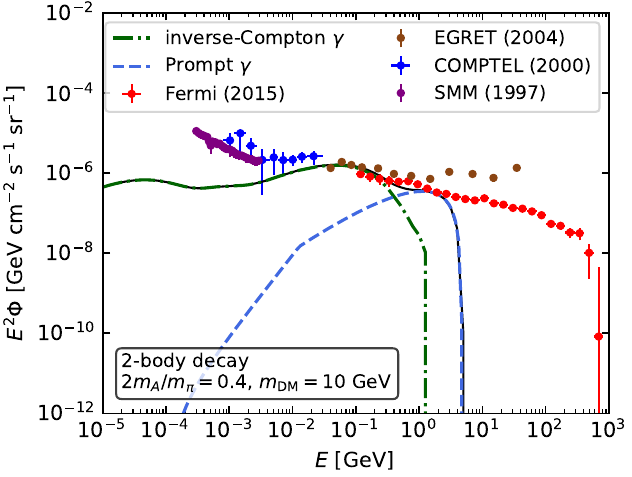}
        \caption{Diffuse $\gamma$-ray background flux from ADM decay is constrained by observed data from various detectors, viz., Fermi-LAT \cite{Fermi-LAT:2014ryh}, EGRET \citep{EGRET:1997qcq, Strong:2004de}, COMPTEL \citep{Weidenspointner_2000} and SMM \citep{Watanabe_2000}. The inverse-Compton emission corresponds to the PD propagation model (see text). The prompt $\gamma$-ray spectrum is that due to the line of sight averaged emission within the Galaxy. Here, the observed data limits the total flux, which corresponds to DM decay lifetime $\tau_{\rm DM}=10^{25}$~s at 95\% C.L.}
    \label{fig:gamma_spectrum}
\end{figure}

\begin{figure*}[t!]
    \centering
    \includegraphics[width=0.48\textwidth]{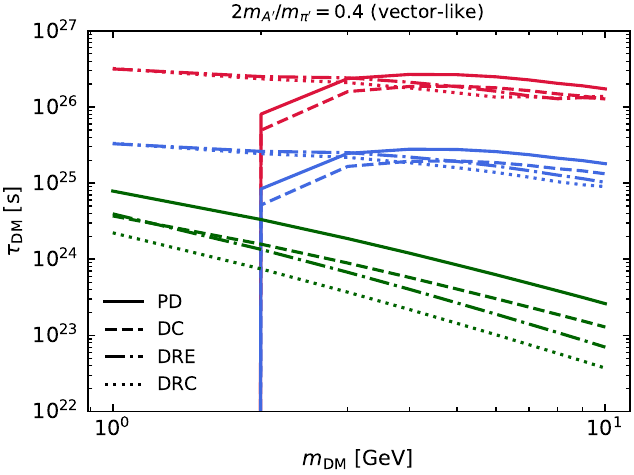}%
    \includegraphics[width=0.48\textwidth]{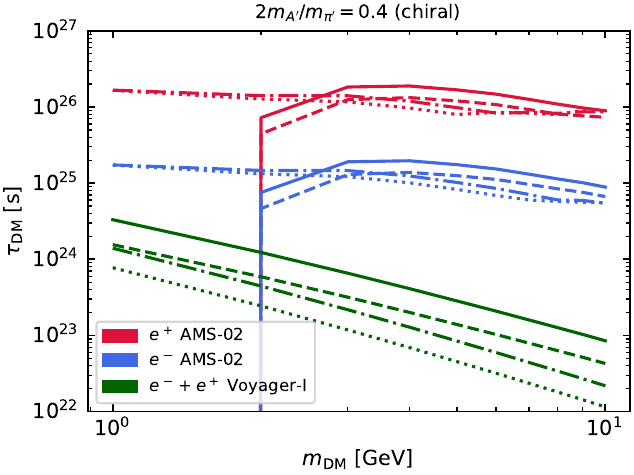}
    \includegraphics[width=0.48\textwidth]{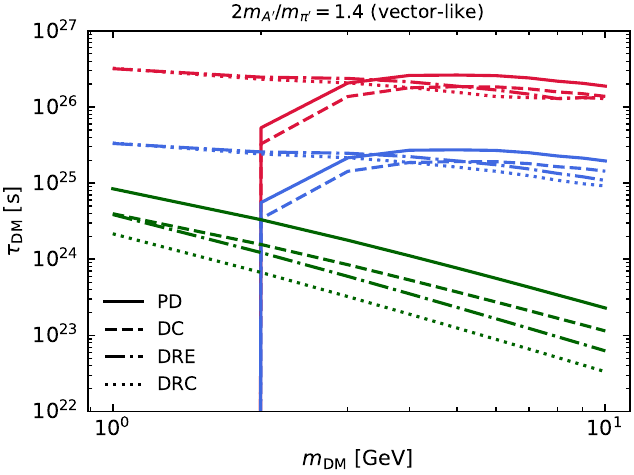}%
    \includegraphics[width=0.48\textwidth]{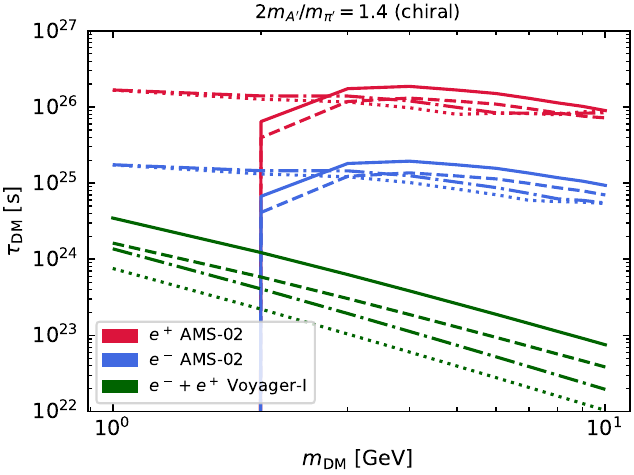}
    \caption{95\% C.L. limits on the ADM lifetime for all the modes considered in this work, including chiral models. The constraints on $e^+$ and $e^-$ are obtained from applying solar modulation to the propagated flux and comparing it to the AMS-02 data. The constraints for $e^+$ and $e^-$ are obtained by comparing the combined unmodulated flux with the Voyager-1 measurements.}
    \label{fig:e_lim}
\end{figure*}

Cosmic-ray (CR) $e^\pm$ diffuse and lose energies in the interstellar medium before reaching Earth. The $e^\pm$s produced from ADM decay constitute an additional CR $e^\pm$ source within the Milky Way. We utilize the publicly available code {\sc GALPROP v57}~\cite{Moskalenko:1997gh,Strong:1998pw,Porter:2021tlr} to solve the CR propagation equation:
\begin{align}
\frac{\partial \psi}{\partial t} =  q(\vec{r}, p)+&\vec{\nabla} \cdot\left(D_{x x} \vec{\nabla} \psi-\vec{V} \psi\right)+\frac{\partial}{\partial p} p^2 D_{p p} \frac{\partial}{\partial p} \frac{\psi}{p^2} \nonumber \\
&-\frac{\partial}{\partial p}\left[\dot{p} \psi-\frac{p}{3}(\vec{\nabla} \cdot \vec{V}) \psi\right],
\end{align}
where $q(\vec{r}, p)$ is the CR source distribution at position $\vec{r}$, $D_{x x}$ is the spatial diffusion coefficient and $D_{p p} \propto D_{x x}$ is the momentum diffusion coefficient. $\vec{V}$ is the convection velocity of the Galactic wind.

Previous studies have attempted to model the propagation of CRs in the Milky Way using local CR measurements~\cite{Porter:2017vaa, Johannesson:2018bit, Johannesson:2019jlk, Silver:2024ero}. In this study, we adopt the propagation models presented in Ref.\cite{Silver:2024ero}, which fit a range of CR species to AMS-02 \cite{AMS:2014xys, Graziani:2017fol} and Voyager data \citep{Stone:2013zlg, Cummings:2016pdr}. Four propagation scenarios are considered as in Ref.\cite{Silver:2024ero}, namely, pure diffusion (PD), diffusion with convection (DC), diffusion with reacceleration (DRE), and diffusion with reacceleration and convection (DRC).

\cref{tab:propagation} presents the best-fit propagation parameters for each scenario described in Ref.~\cite{Silver:2024ero}. The spatial diffusion coefficient $D_{xx} \propto \beta^\eta \rho^\delta$, where $\beta = v/c$ and $\rho$ denotes the rigidity. $D_{xx}$ is modeled by a broken power-law function of $\rho$ and is normalized by $D_{0, xx}$ at $\rho = 4$ GV. For $\rho < \rho_\mathrm{break}$, $\delta = \delta_1$, and for $\rho > \rho_\mathrm{break}$, $\delta = \delta_2$. $v_A$ represents the Alfvén velocity that governs reacceleration and determines $D_{p p}$~\cite{Moskalenko:1997gh}. $dV/dz$ denotes the gradient of the convection velocity along the Galactic plane.
$\Phi$ is the solar modulation potential in force-field approximation~\cite{1968ApJ...154.1011G}.
\begin{table}[b]
    \caption{CR propagation parameters taken from Ref.~\cite{Silver:2024ero}.}
    \begin{ruledtabular}
    \begin{tabular}{ccccc}
      & PD & DC & DRE & DRC \\\colrule
    \makecell{$D_{0, x x}$ \\ $[10^{28}$ cm$^{2}$ s$^{-1}]$}  & 4.5767 & 3.6183 & 4.7776 & 4.4452 \\
    $\delta_1$  & 0.4047 & 0.4448 & 0.4052 & 0.4163 \\
    $\delta_2$  & 0.1928 & 0.1975 & 0.2315 & 0.2404 \\
    $\rho_\mathrm{break}$ [GV] & 290.67 & 283.29 & 308.04 & 308.04 \\
    $\eta$  & 0.0004 & 0.8196 & 0.3851 & 0.4373 \\
    $v_A$ [km s$^{-1}$]  & ... & ... & 26.727 & 32.187 \\
    \makecell{$dV/dz$ \\ $[\mathrm{km} \mathrm{s}^{-1} \mathrm{kpc}^{-1}]$}  & ... & 10.022 & ... & 6.3482 \\
    $\Phi$ [MV] & 368 & 375 & 612 & 622 \\
    \end{tabular}
    \end{ruledtabular}
    \label{tab:propagation}
\end{table}

We follow Ref.~\cite{Song:2019nrx} and modify the \texttt{gen\_DM\_source.cc} routine provided by {\sc GALPROP} to include the $e^\pm$ spectra resulting from the decay of composite ADM. This routine enables the user to define additional source terms $q(\vec{r}, p)$ arising from various dark matter spatial distributions and injection spectra. Previous applications of this routine include studies of inverse Compton emission from millisecond pulsars in the Galactic center~\cite{Song:2019nrx, Macias:2021boz}, as well as dark matter annihilation in M31~\cite{Egorov:2022fkc}.

We propagate $e^\pm$ fluxes using the parameters in \cref{tab:propagation} and compare the predicted fluxes at Earth with AMS-02~\cite{AMS:2014xys, Graziani:2017fol} and Voyager data \citep{Stone:2013zlg, Cummings:2016pdr}. It is noteworthy that the propagation models derived in Ref.~\cite{Silver:2024ero} do not explicitly fit the local CR $e^\pm$ data. Nonetheless, they provide reasonable estimates for the propagation of $e^\pm$ in various scenarios. \cref{fig:epm_spectrum} shows the $e^\pm$ spectra from ADM for four propagation scenarios. We present the case corresponding to two-body decay of the dark meson in the vector-like model with $m_{\rm DM}=10$~GeV and $2 m_{A'} / m_{\pi'}=0.4$, assuming a fiducial value of $\tau_\mathrm{DM} = 10^{26}$~s. The red lines represent the predicted spectra outside of the solar hemisphere (without solar modulation), while the green lines show the predicted spectra at Earth (with solar modulation). We also present the AMS-02 (blue for $e^-$ and orange for $e^+$) and Voyager $e^\pm$ (black) data.

Additionally, we calculate the corresponding inverse-Compton emission by $e^\pm$ arising from ADM decay, utilizing the standard 2D interstellar radiation field in GALPROP. We use the all-sky average flux of inverse-Compton emission to put constraints on composite ADM lifetime. \cref{fig:gamma_spectrum} shows the propagated $\gamma$-ray spectrum for $m_{\rm DM}=10$~GeV and $2 m_{A'} / m_{\pi'}=0.4$ for the vector-like model. The total flux (black line) arises from the Galactic $\gamma$-ray component (blue dashed line), originating from the prompt DM decay $\gamma$-ray flux and the inverse-Compton scattering of the interstellar radiation field by high-energy $e^\pm$s (green dashed-dotted line) for the DRC model. We find that the primary $\gamma$-ray spectrum from ADM decay is subdominant in constraining the ADM decay, although it dominates at the high-energy end of the total $\gamma$-ray flux for higher $m_{\rm DM}$ values considered in this study.

\subsection{Constraints by $e^\pm$s and $\gamma$ rays}

We consider $e^\pm$ observations from AMS-02~\cite{AMS:2014xys, Graziani:2017fol} and Voyager~\citep{Stone:2013zlg, Cummings:2016pdr}. Fermi-LAT measures the diffuse isotropic $\gamma$-ray background (IGRB) between 100 MeV and 820 GeV~\cite{Fermi-LAT:2014ryh}. We also use soft $\gamma$-ray data from EGRET \citep{EGRET:1997qcq, Strong:2004de}, COMPTEL \citep{Weidenspointner_2000}, and SMM \citep{Watanabe_2000} to constrain the ADM decay lifetime at lower masses.
The lower limits on the ADM decay lifetime $\tau_{\rm DM}$ are evaluated by the condition that the expected astrophysical flux in any energy bin $i$ satisfies $J_i\leqslant M_i + n\times \Sigma_i$, where $M_i$ is the observed astrophysical flux and $\Sigma_i$ is the error in the $i$-th energy bin~\cite{Fermi-LAT:2010qeq}. The values $n=1.28$, 1.64, and 4.3 correspond to 90\%, 95\%, and 99.9999\% confidence level (C.L.) lower limits, respectively.

\cref{fig:e_lim} shows the 95\% C.L. constraints on $\tau_{\rm DM}$ from $e^\pm$, obtained individually and separately using the AMS-02 and Voyager data, respectively. For models without reacceleration, the predicted ADM $e^\pm$ spectra for $m_{\rm DM} \lesssim 2.4$ GeV, after accounting for solar modulation, remain below the threshold of the lowest energy data point observed by AMS-02. Consequently, limits for $m_{\rm DM} \lesssim 2.4$ GeV cannot be provided by AMS-02 observations when reacceleration is not considered\footnote{We note that the lower limit on $m_{\rm DM}$ that can be probed by AMS-02 data also depends on the assumed solar modulation potential. The lower limit will increase for a larger potential as the maximum $e^\pm$ energy after solar modulation decreases, and vice versa.}. This is indicated by the sharp drop in \cref{fig:e_lim}. However, for the models with reacceleration, limits can be obtained over the entire mass range considered. We find that the most stringent limits are obtained from the positron flux for the two-body decay of the dark meson in the vector-like model, with $2 m_{A'} / m_{\pi'}=0.4$, corresponding to approximately $3\times10^{27}$~s at $m_{\rm DM}=1$ GeV. The constraints obtained for the chiral models are weaker than the others.

Voyager measures the cosmic-ray spectrum outside of the solar hemisphere in the local interstellar medium and hence is unaffected by solar modulation \citep{Stone:2013zlg, Cummings:2016pdr}. We use the combined $e^+/e^-$ spectrum data observed by Voyager to constrain the ADM decay lifetime. This is shown for two- and three-body decay of the dark meson and for both vector and chiral models in \cref{fig:e_lim}. Additionally, we present the constraints for various propagation models considered. The models without reacceleration (PD and DC) lead to more stringent limits at higher energies. At approximately 10~GeV, the limits obtained range between $3\times10^{22}$~s for the DRC model to $2\times10^{23}$~s for the PD model. At around 1 GeV, the constraints obtained from the DC and DRE models coincide and correspond to approximately $3\times10^{24}$~s, while for various propagation models, the limits range between $2$-$5\times 10^{24}$~s.

\begin{figure}[t!]
    \centering
    \includegraphics[width=0.45\textwidth]{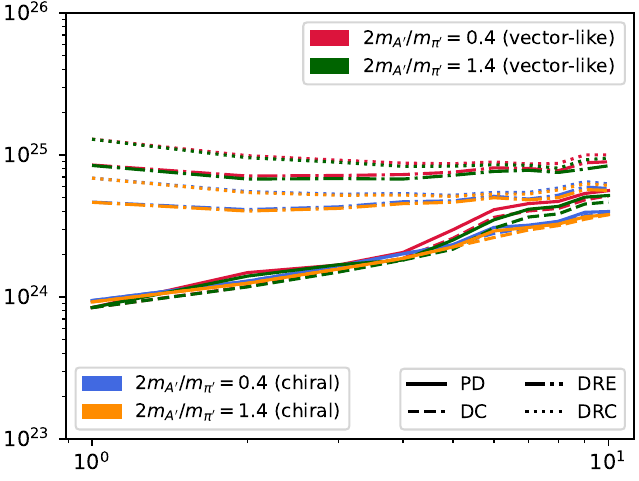}%
    \caption{95\% C.L. limits on the ADM decay lifetime for all the decay modes considered in this work for $\gamma$ rays.}
    \label{fig:gamma_lim}
\end{figure}

\cref{fig:gamma_lim} presents our constraints for various propagation and ADM decay models using $\gamma$-ray measurements. At lower energies, the DRC model (which considers reacceleration) provides the most stringent constraint, $\tau_{\rm DM}\gtrsim 1.5\times 10^{25}$~s at $m_{\rm DM}=1$ GeV. The constraints obtained for the PD and DC models are less restrictive, which are as low as $\gtrsim 7\times10^{23}$~s at $m_{\rm DM}=1$~GeV for the PD propagation model. At $m_{\rm DM}=10$ GeV, the limits range between $\tau_{\rm DM}\gtrsim 3\times10^{24}$~s and $1.2\times10^{25}$~s. Thus, the constraints obtained from $\gamma$ rays are comparable to those derived from $e^-$s using AMS-02 data. The limits for a given propagation model are weaker for the two-body decay of the dark meson in chiral composite ADM than the other cases.

\subsection{Neutrino constraints}

Neutrinos are weakly interacting particles and are undeflected by cosmic magnetic fields. They can reach Earth from sufficiently high redshifts. The neutrino spectrum from ADM decay is monoenergetic, as explained in \cref{sec:decay}. The prompt spectrum from dark meson decay is equally distributed among $\nu_e$ $\nu_\mu$, and $\nu_\tau$. However, the cross section of $\nu_\tau$ events in the currently operating water Cherenkov detectors, like Super-K \cite{Super-Kamiokande:2002weg}, is diminutive \citep{Super-Kamiokande:2017edb}. Hence, for our analysis, we consider only $\nu_e$ and $\nu_\mu$ events at Earth and compare them with the atmospheric background model of these flavors derived from observations by the Super-K detector \citep{Super-Kamiokande:2015qek}.
\begin{figure*}
    \centering
    \includegraphics[width=0.45\textwidth]{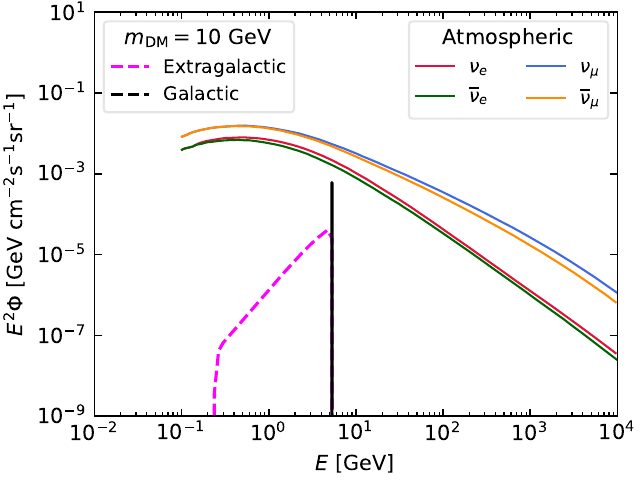}%
    \includegraphics[width=0.45\textwidth]{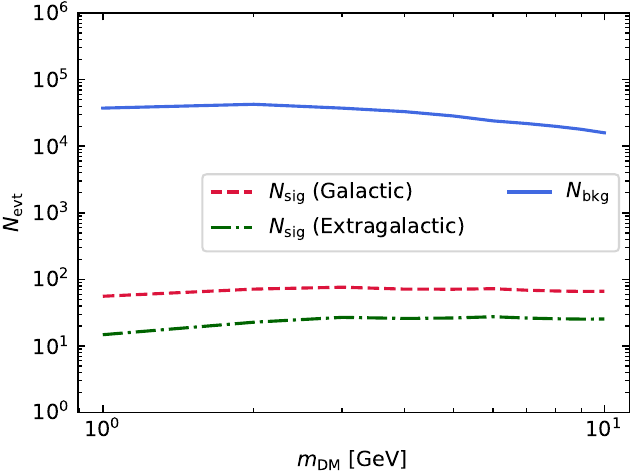}
    \caption{\textit{Left}: Sum of the fluxes of $\nu_e$, $\nu_\mu$, and their antineutrinos at Earth for $m_{\rm DM}=10$~GeV, where the two body decay case with $2 m_{A'} / m_{\pi'}=0.4$ for the vector-like model is assumed. Neutrino mixing is taken into account.
    The individual components of the neutrino flux, viz., Galactic (black) and extragalactic (brown) fluxes, are shown for a fiducial value of DM decay lifetime $\tau_{\rm DM}=10^{25}$~s. The prompt neutrino spectrum from ADM decay is monoenergetic, which shows up as a spike in the Galactic component (see text). We show the atmospheric neutrino flux for individual flavors, obtained from Super-K observations \citep{Honda:2015fha}. \textit{Right}: The number of events expected at the Super-K site from the atmospheric background and ADM decay signal, assuming both FC and PC events. Here, we use a fiducial value of $\tau_{\rm DM}=10^{25}$~s to calculate the number of signal events. The Galactic and extragalactic components are shown separately. }
    \label{fig:neu}
\end{figure*}

Super-K has sensitivities to not only MeV neutrinos but also neutrinos with $E_\nu>100$ MeV or higher \cite{Super-Kamiokande:2002weg}. Neutrino events in Super-K can be classified into fully contained (FC), partially contained (PC), and upward-going muon (UPMU) events. The FC and PC events have a reconstructed neutrino interaction vertex inside the fiducial volume of the inner detector, which is optically separated from the outer detector. They are distinguished by the number of triggered photomultiplier tubes in the outer detector. The typical neutrino energy for FC (PC) events is between 0.1 and 10 GeV (1 and 100 GeV), and the detector is sensitive to $\nu_e$, $\nu_\mu$, $\overline{\nu}_e$, $\overline{\nu}_\mu$. We use the effective areas from the analysis presented in Ref.~\cite{Super-Kamiokande:2021dav} and the associated data release for different samples (FC or PC) and flavors. 
Since we are primarily interested in neutrinos at multi-GeV energies or lower, we do not include UPMU events.
As an approximation, we average the effective area $A_{\rm eff}^{(s,f)}(E_\nu, \Omega)$ for a given sample $s$ and flavor $f$ over the source direction. 
We focus on $\nu_e$ and $\nu_\mu$ and their antineutrinos at Earth after taking into account neutrino mixing in the analysis.

The atmospheric neutrinos provide the primary background for neutrino detection at $0.1-10$~GeV energies. For our analysis, we use the atmospheric neutrino flux given in Ref.~\citep{Honda:2015fha}, which is based on the Honda, Kasahara, Kajiyama, and Mihara (also called HKKM) model \citep{Honda:2004yz, Honda:2006qj, Honda:2011nf}. The left panel of \cref{fig:neu} shows the all-direction averaged atmospheric neutrino flux model calculated in Super-K for $\nu_e$, $\nu_\mu$, and their antineutrinos. The total number of background events in a specific energy bin is obtained by summing over all flavors and taking into account both FC and PC events for Super-K.
The mean number of events from the atmospheric background flux model is estimated by
\begin{equation}
    N_{\rm bkg}^{(s,f)} = T_{\rm ob} \int dE_\nu^r \int dE_\nu d\Omega\dfrac{dN^{(s,f)}}{dE_\nu} A_{\rm eff}^{(s,f)}  G(E_\nu^r, E_\nu)
    \label{eqn:bkgd}
\end{equation}
Here $T_{\rm ob}$ is the observation time for the Super-K detector. We use the time period from the SK-I to the SK-V phase, corresponding to the period between April 1996 and July 2020, which is 24.35 years. The reconstructed neutrino energy $E_\nu^r$ depends on the intrinsic property of the detector and varies from the actual neutrino energy $E_\nu$. The energy resolution function $G(E_\nu^r, E_\nu)=\exp(-(E_\nu^r - E_\nu)^2/2\sigma_E^2)/\sqrt{2\pi}\sigma_E$ is a normalized Gaussian distribution with a width of $\sigma_E/E_\nu=0.1$ corresponding to $\sim10-20\%$ energy resolution for GeV and higher energies, and $\Delta \log_{10}E_\nu=0.2$ is used in this work
~\cite{Super-Kamiokande:2015qek}.

In the left panel of \cref{fig:neu}, we show the sum of $\nu_e$ and $\nu_\mu$ neutrino fluxes at Earth, for $m_{\rm DM}=10$~GeV, and a fiducial value of DM decay lifetime $\tau_{\rm DM}=10^{25}$~s. We consider the flavor oscillation during propagation over cosmological distances, using the following expression,
\begin{equation}
    {\mathcal{Q}}_{\nu_\alpha}^{\oplus} = \sum_{j=1,2,3} |U_{\alpha j}|^2 |U_{\beta j}|^2 {\mathcal{Q}}_{\nu_\beta}^{\rm source},
\label{FlavorToMassConversion}
\end{equation}
where $\alpha, \beta=e, \mu, \tau$ and $U$ is the Pontecorvo-Maki-Nakagawa-Sakata (PMNS) matrix. ${\mathcal{Q}}_{\nu_\alpha}$ is the flux of neutrinos of flavor $\alpha$. We use the best-fit parameter values from {\tt NuFIT 2022} \cite{Esteban2020} to account for neutrino flavor mixing. The Galactic component (black solid line) exhibits line emission, while the extragalactic component (magenta dashed line) is smoothed out due to the integration over redshifts (cf. \cref{eqn:extragalactic}). We consider a maximum redshift of $z=20$ in this study for estimating the limits on $\tau_{\rm DM}$. It can be seen that the dominant contribution arises from the prompt Galactic spectrum. Energy losses due to the adiabatic expansion of the universe are also included. The right panel in \cref{fig:neu} shows the expected number of background events and signal events from ADM decay at the Super-K site, corresponding to an observation time of SK-I to SK-V phase and the same fiducial value of $\tau_{\rm DM}=10^{25}$~s. 

We obtain the upper limit on the DM lifetime by adopting an analogous approach to that done in Refs.~\citep{Super-Kamiokande:2002hei, Beacom:2006tt, Yuksel:2007ac, Palomares-Ruiz:2007trf}. 
We stress that our analysis is suitable for obtaining conservative bounds, and more sophisticated analyses with actual Super-K data are encouraged. 
We consider three energy bins of width $\Delta \log_{10}E_\nu=0.2$ for each value of $m_{\rm DM}$, such that the central bin includes the galactic flux component, which has the dominant contribution in our case. Hence, the total energy range for our neutrino analysis is $10^{-0.3}E_0$ to $10^{+0.3}E_0$ divided into three logarithmic bins, where $E_0$ is the energy at which the Galactic line emission appears for each $m_{\rm DM}$. We choose a conservative value for the bin width since the DM signal is sharply peaked at $E_\nu \approx 0.5 m_{\rm DM}$. The $\chi^2$ function is defined as a sum over energy bins,
\begin{align}
    \chi^2 = \sum_{l=1}^3 \dfrac{(\alpha A_l +\beta B_l - N_l)^2}{\sigma_{\rm stat}^2 + \sigma_{\rm sys}^2}
\end{align}
where $A_l$ and $B_l$ are the fractions of DM decay neutrino events and atmospheric background events in the $l$-th bin, compared to the total number of corresponding events summed over all bins~\citep[see, e.g.,][]{Palomares-Ruiz:2007trf}. Here $N_l$ is the number of background events in the $l$-th bin obtained using \cref{eqn:bkgd}, and $B_l=N_l/N_{\rm bkg}$.
The fraction $A_l$, which is for the signal part, is similarly calculated by propagating the neutrino spectrum obtained in our composite ADM model, using \cref{eqn:los,eqn:extragalactic} by integrating over energy bins. In our analysis, we consider the systematic uncertainty to be $15\%$, i.e, $\sigma_{\rm sys}=0.15N_l$ \citep{Barr:2006it, Evans:2016obt} and 1$\sigma$ bound on the statistical error, i.e., $\sigma_{\rm stat}^2=N_l$. The free parameters $\alpha$ and $\beta$ determine the strength of signal and background events, respectively. The value of $\tau_{\rm DM}$ is determined from the strength of the DM signal obtained by best-fit values of $\alpha$. Since the number of background events is $\gg1$, we treat $\alpha$  as a discrete parameter. When both parameters are varied simultaneously, the best fit for minimizing the $\chi^2$ corresponds to $\alpha=0$, indicating no DM neutrino signal in the observed flux by Super-K. The minimum value of $\chi^2$ for a specific value of $\alpha$ corresponds to a probability value $P(\alpha)\propto\exp(-\chi^2(\alpha))$. In this work, it is sufficient to treat $\alpha$ as an integer, and we vary $\alpha=1,2,3,...$ and find the best-fit value of $\beta$ for each $\alpha$, to yield a probability distribution normalized to unity. We exclude the DM hypothesis at 95\% C.L. by calculating the value of $\alpha$ for which the $p$-value is equal to 0.05. The number of DM events corresponding to the upper limit is then simply $n_{\rm DM}^{\rm UL}=\sum_l\alpha_{95} A_l$.

\begin{figure}[t!]
    \centering
    \includegraphics[width=0.45\textwidth]{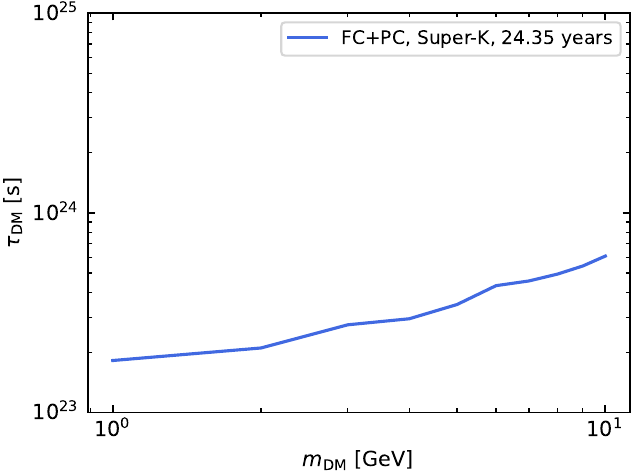}
    \caption{95\% C.L. limits on DM decay lifetime from Super-K. Note that both vector-like and chiral models lead to the same neutrino constraint.  
    }
    \label{fig:neu_lim}
\end{figure}

\begin{figure}[t!]
    \centering
    \includegraphics[width=0.48\textwidth]{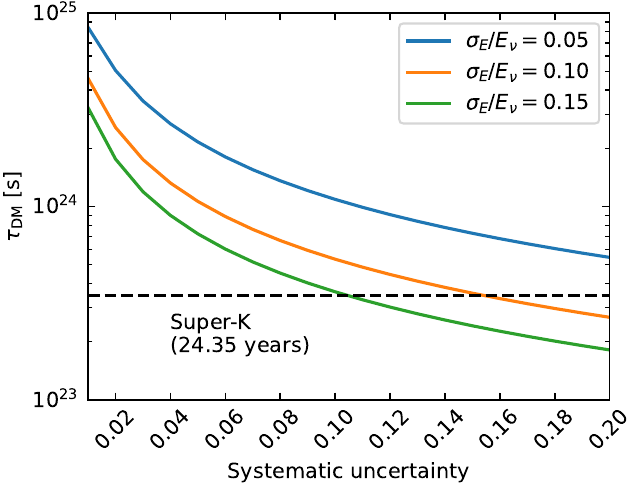}
    \caption{Dependence of projected constraints for Hyper-K on $\sigma_\mathrm{sys}$ and $\sigma_E/E_\nu$, considering $m_\mathrm{DM} = 5$~GeV and an observation time of $T_{\rm obs} = 10$ years. The dashed horizontal line shows the constraint obtained from Super-K, where $\Delta \log_{10}E_\nu=0.2$ and $\sigma_\mathrm{sys}/N_l=15$\% are assumed.}
    \label{fig:nusys}
\end{figure}

The corresponding value of $\tau_{\rm DM}$ is derived from the upper limit on $n_{\rm DM}$ for each specific value of $m_{\rm DM}$. This is shown in \cref{fig:neu_lim}. 
During the 24.35-year observation period of SK-I, we find that at lower energies, the constraints correspond to $\tau_{\rm DM}\gtrsim 2\times 10^{23}$~s, whereas at 10~GeV, the constraints can reach $\gtrsim 6\times10^{23}$~s.
Hyper-K, the successor to Super-K, is currently under construction~\cite{Abe:2011ts}.
The event rates in Hyper-K can be estimated by scaling the effective detection area of Super-K by the ratio of fiducial masses, which is approximately $8.3$ (187~kton for Hyper-K vs. 22.5~kton for Super-K). 
However, the neutrino constraints from the forthcoming Hyper-K observatory are not expected to scale directly with the fiducial mass ratio due to the dominance of systematic uncertainties from the measurement of atmospheric neutrinos. In \cref{fig:nusys}, assuming $\Delta \log_{10}E_\nu=2\sigma_E/E_\nu$, we illustrate the dependence of the constraints from Hyper-K on $\sigma_\mathrm{sys}$ and $\sigma_E/E_\nu$. For this analysis, we consider the case of $m_\mathrm{DM} = 5$~GeV and an observation time of $T_{\rm obs} = 10$ years for Hyper-K. Only FC events are utilized, as the scaling of the effective area is nontrivial for PC events. The dashed horizontal line in \cref{fig:nusys} denotes the constraint obtained from Super-K using the fiducial parameters. 
At a $\sim10$\% energy resolution, the constraints from Hyper-K would be only comparable to those of Super-K, provided that $\sigma_\mathrm{sys}/N_l$ remains around 15\%. To achieve substantial improvements in the constraints, which is especially useful for overcoming the CMB constraint below 2~GeV, understanding the atmospheric neutrino background with systematic uncertainty at a level of $\lesssim10$\% is required. Alternatively, the constraints can be improved with better energy resolution. For instance, a resolution better than the 10\% level may provide clear improvements over Super-K, and such a level of precision might be achievable in experiments such as DUNE~\cite{Friedland:2018vry}.

\section{\label{sec:conclusion}Summary}
We presented comprehensive studies on the composite ADM.
We explored four models for ADM decay. 
We considered two values of the mass ratio between dark photon and dark meson ($2m_A'/m_\pi'$) that provide different dominant channel of the dark meson decay, where each can have either two-flavor vector-like dark quarks or three-flavor chiral dark quarks. 

We investigated multimessenger signatures arising from ADM decay, viz., $e^+$, $e^-$, $\gamma$ rays, and $\nu$-s. 
We performed a detailed analysis of the constraints on the ADM decay since the ADM mass lies in the range of $1$--$10$~GeV in the composite ADM scenario.
The decay products $e^\pm$ and $\gamma$ from the ADM with the low mass are less energetic, which can be out of the reach of AMS-02 and Fermi-LAT. 
Therefore, we also considered the data from other observations sensitive to the low-energy $e^\pm$ and $\gamma$-ray fluxes. 
As for the constraint from the neutrino signals, we put a conservative bound on the ADM lifetime by the approach with the $\chi^2$ function. 

While the constraints from $e^+/e^-$ depend on the specific propagation model employed, they yield the most stringent limits corresponding to $\gtrsim10^{26}$~s for all ADM models. 
The limits obtained from $\gamma$ rays range between $10^{24}-10^{25}$~s, and those obtained from neutrino observations with Super-K are also comparable. 
We showed that neutrino observations are useful for obtaining robust constraints, and future neutrino observations with Hyper-K and DUNE can improve the bounds. 
In addition, the IceCube-Gen2 upgrade~\citep{IceCube-Gen2:2020qha} and the ORCA instrument of KM3Net~\cite{Miranda:2022kzs} will also be sensitive to neutrino detection above $\sim 1$--$3$~GeV.

\section*{Acknowledgements}

A.K. and T.K. thank Keiichi Watanabe for his collaboration during the early stages of this work. S.D., K.M., and D.S. thank Arman Esmaili for discussions on Hyper-K analyses.  
A.K. acknowledges partial support from the Norwegian Financial Mechanism for years 2014-2021, grant nr 2019/34/H/ST2/00707; and from the National Science Centre, Poland, grant DEC-2018/31/B/ST2/02283. 
The work of T. K. is supported in part by the National Science Foundation of China under Grant Nos.~11675002, 11635001, 11725520, 12235001, and 12250410248. This work is supported by NSF Grants Nos.~AST-2108466 (K.M.), AST-2108467 (K.M.), AST-2308021 (K.M.), and KAKENHI No.~20H05852 (S.D., K.M., and D.S.). 

\appendix 
\section{Dark Baryon Decay Rates \label{app:DR}}

Dark baryons decay into dark mesons and antineutrinos through portal interactions given in \cref{eq:Intermediate_Portal,eq:Intermediate_Portal_Ch}.
The dark mesons decay into dark photons only through a chiral anomaly in the dark sector, and hence, most dark baryons decay into the darkly-charged mesons that do not decay into dark photons and leave only the antineutrino signals.
The branching ratios into the dark-neutral mesons leading to the cosmic-ray signals are significant for our purpose.

First, the decay matrix elements can be computed by the Clebsch-Gordan coefficients (CGCs) and a representative element.
The labels of states $|j,m\rangle$, and decay operators $\mathcal{O}_M^{(J)}$ denote the flavor index.
The decay amplitudes are represented as 
\begin{align}
	\langle j,m|\mathcal{O}_M^{(J)}|j',m'\rangle 
	= \langle j',m';J,M|j,m\rangle 
	\langle{j}|\mathcal{O}^{(J)}|j'\rangle \,.
\end{align}
Here, $|j',m'\rangle$ denotes the initial dark baryon state, while $\langle j,m|$ denotes the final dark meson state. 
$\langle j',m';J,M|j,m\rangle $ is the Clebsch-Gordan coefficient.
$\langle{j}|\mathcal{O}^{(J)}|j'\rangle $ is referred to as a reduced matrix element, which is independent of the quantum numbers $m \,, m'$, and $M$.

Let us consider the composite ADM with $N_f = 2$ vector-like dark quarks. 
There is one dark-neutral operator in terms of the group structure of the flavor symmetry, denoted by $\mathcal{O}_{n'} = U'D'D'$.
This operator leads to two dark baryon decay processes, $p' \to \pi'^+ + \overline \nu$ and $n' \to \pi'^0 + \overline \nu$, see \cref{eq:meson-baryon_ADM} in the text.
The baryon states and the baryon operators are in the fundamental representation of $SU(2)$, while the meson states are in the adjoint representation of $SU(2)$.
Since the irreducible decomposition of $\mathbf{2} \otimes \mathbf{2}$ includes a single $\mathbf{3}$ representation, the decay amplitudes are parameterized by a single reduced matrix $A_n$ and are related to each other by the CGCs.
\begin{align}
	\langle \pi'^+|\mathcal{O}_{n'}|p'\rangle & = A_{n} \,, \\
	\langle \pi'^0|\mathcal{O}_{n'}|n'\rangle & = \frac{1}{\sqrt{2}} A_{n} \,.
\end{align}
Therefore, the fraction decaying into $\pi'^0$ is independent of the model parameters. 
\begin{align}
	\frac{\Gamma(n' \to \pi'^0 \overline \nu)}{\Gamma(n' \to \pi'^0 \overline \nu) + \Gamma(p' \to \pi'^+ \overline \nu)} = \frac13 \,.
\end{align}

Next, let us consider the composite ADM with $N_f = 3$ chiral dark quarks. 
There are two dark-neutral operators consisting of $U' \,, D' \,, S'$: one is an operator $\mathcal{O}_{\Lambda'}$ transforming as $\Lambda'$ (singlet under the isospin-like subgroup ``dark isospin'' of flavor symmetry), and another is an operator $\mathcal{O}_{\Sigma'}$ transforming as $\Sigma'_3$ (a neutral component of a triplet of the dark isospin).
Contrast to the case with $N_f = 2$, the irreducible decomposition of $\mathbf{8} \otimes \mathbf{8}$ includes two $\mathbf{8}$ representations. 
The decay amplitudes are parameterized by two reduced matrix elements $A^{(1)}$ and $A^{(2)}$ and are related to each other by the CGCs of $SU(3)$. 
One can find the CGCs of $SU(3)$ in Ref.~\cite{McNamee:1964xq}. 
The matrix elements for the operator $\mathcal{O}_{\Lambda'}$ are 
\begin{align}
	\langle \pi'_3|\mathcal{O}_{\Lambda'}|\Sigma'_3 \rangle & = \frac{1}{\sqrt{5}} A^{(1)} \,, \\
	\langle \pi'|\mathcal{O}_{\Lambda'}|\Sigma' \rangle & = \frac{1}{\sqrt{5}}A^{(1)} \,, \\
	\langle \pi'^\dag|\mathcal{O}_{\Lambda'}|\Sigma'^\dag \rangle & = \frac{1}{\sqrt{5}}A^{(1)} \,, \\
	\langle \eta'|\mathcal{O}_{\Lambda'}|\Lambda'\rangle & = - \frac{1}{\sqrt{5}} A^{(1)} \,, \\
	\langle K'_1|\mathcal{O}_{\Lambda'}|p'\rangle & = - \frac{1}{2\sqrt{5}} A^{(1)} + \frac12 A^{(2)} \,, \\
	\langle K'_2|\mathcal{O}_{\Lambda'}|n'\rangle & = - \frac{1}{2\sqrt{5}} A^{(1)} + \frac12 A^{(2)} \,, \\
	\langle K'^\dag_1|\mathcal{O}_{\Lambda'}|\Xi'_1\rangle & = - \frac{1}{2\sqrt{5}} A^{(1)} - \frac12 A^{(2)} \,, \\
	\langle K'^\dag_2|\mathcal{O}_{\Lambda'}|\Xi'_2\rangle & = - \frac{1}{2\sqrt{5}} A^{(1)} - \frac12 A^{(2)} \,. 
\end{align}
On the other hand, the matrix elements for the operator $\mathcal{O}_{\Sigma'}$ are 
\begin{align}
	\langle \eta'|\mathcal{O}_{\Sigma'}|\Sigma'_3 \rangle & = \frac{1}{\sqrt{5}} A^{(1)} \,, \\
	\langle \pi'|\mathcal{O}_{\Sigma'}|\Sigma' \rangle & = \frac{1}{\sqrt{3}} A^{(2)} \,, \\
	\langle \pi'^\dag|\mathcal{O}_{\Sigma'}|\Sigma'^\dag \rangle & = - \frac{1}{\sqrt{3}} A^{(2)} \,, \\
	\langle \pi'_3|\mathcal{O}_{\Sigma'}|\Lambda'\rangle & = \frac{1}{\sqrt{5}} A^{(1)} \,, \\
	\langle K'_1|\mathcal{O}_{\Sigma'}|p'\rangle & = \frac{1}{2} \sqrt{\frac{3}{5}} A^{(1)} + \frac{1}{2\sqrt{3}} A^{(2)} \,, \\
	\langle K'_2|\mathcal{O}_{\Sigma'}|n'\rangle & = - \frac{1}{2} \sqrt{\frac{3}{5}}  A^{(1)} - \frac{1}{2\sqrt{3}} A^{(2)} \,, \\
	\langle K'^\dag_1|\mathcal{O}_{\Sigma'}|\Xi'_1\rangle & = \frac{1}{2} \sqrt{\frac{3}{5}} A^{(1)} - \frac{1}{2\sqrt{3}} A^{(2)} \,, \\
	\langle K'^\dag_2|\mathcal{O}_{\Sigma'}|\Xi'_2\rangle & = - \frac{1}{2} \sqrt{\frac{3}{5}} A^{(1)} + \frac{1}{2\sqrt{3}} A^{(2)} \,.
\end{align}
We give the convention for the names of baryon and meson in the chiral composite ADM in \cref{eq:meson-baryon_CADM}. 
We define the amplitudes normalized by the Wilson coefficient, $\widetilde A^{(i)}_{N} = M^{-3}_{N} A^{(i)}$
The fractions decaying into $\pi'_3$ and $\eta'$ are given by
\begin{align}
	& \frac{\Gamma(\Sigma_3' \to \pi'_3 \overline \nu) + \Gamma(\Lambda' \to \pi'_3 \overline \nu)}{\Gamma_\mathrm{total}} \nonumber \\
	& = 
	\frac{\Gamma(\Sigma_3' \to \eta' \overline \nu) + \Gamma(\Lambda' \to \eta' \overline \nu)}{\Gamma_\mathrm{total}} \nonumber \\
	& = \frac{1}{5} \frac{|\widetilde A^{(1)}_{\Lambda}|^2+|\widetilde A^{(1)}_{\Sigma}|^2}{|\widetilde A^{(1)}_{\Lambda}|^2 + |\widetilde A^{(2)}_{\Lambda}|^2 + |\widetilde A^{(1)}_{\Sigma}|^2 + |\widetilde A^{(2)}_{\Sigma}|^2} \,.
\end{align}
Here, $\Gamma_\mathrm{total}$ denotes the sum of decay widths of all dark baryons induced by the operators $\mathcal{O}_{\Lambda'}$ and $\mathcal{O}_{\Sigma'}$.
We take $\widetilde A^{(1)}_{\Lambda}:\widetilde A^{(2)}_{\Lambda}:\widetilde A^{(1)}_{\Sigma}:\widetilde A^{(2)}_{\Sigma} = 1:1:1:1$ in our numerical simulations. 

\section{Cascade Decay \label{app:cascade}}

Following Refs.~\cite{Mardon:2009rc,Elor:2015tva}, we compute the energy spectra of decay products through the cascade decay of a particle.
Now we consider the decay chain $\phi_{n+1} \to \phi_n + X_n \to \cdots \to \psi + X$.
$\phi_{n+1}\,,\phi_n\,,\cdots$ denote particles producing a decay product $\psi$ of our interest, and $X$ collectively denotes the other products. 
Once we have an energy spectrum of $\psi$ in the rest frame of $\phi_n$, we obtain that in the rest frame of $\phi_{n+1}$ by boosting the system. 
The velocity of $\phi_n$ in the rest frame of $\phi_{n+1}$ is 
\begin{align}
	|\beta_{n+1}| & = \frac{1}{1 + \epsilon_{n+1} \eta_{n+1}} \sqrt{\left( 1 - \epsilon_{n+1}^2 \right)\left( 1 - \eta_{n+1}^2 \right)} \,, \nonumber \\
	\epsilon_{n+1} & = \frac{m_n+m_{X_n}}{m_{n+1}} \,, \quad 
	\eta_{n+1} = \frac{m_n-m_{X_n}}{m_{n+1}} \,.
\end{align}
Here, $m_n$ denotes the mass of $\phi_n$, and $m_{X_{n}}$ denotes the invariant mass of the collective products $X_{n}$.
We obtain the energy of $\psi$ in the rest frame of $\phi_{n+1}$ by boosting that in the rest frame of $\phi_n$. 
$E_i$ denotes the energy of $\psi$ in the rest frame of $\phi_i$, and its relation is given by $E_{n+1} = \gamma_{n+1} (E_n - \beta_{n+1} p_n^\parallel)$.
Here, $p_n^\parallel = \sqrt{E_n^2 - m_\psi^2} \cos\theta_n$ is the momentum of $\psi$ in the rest frame of $\phi_{n+1}$, and $\theta_n$ is the angle of $\psi$ relative to the boost axis.
Defining the dimensionless parameters, $x_i \equiv 2 E_i/m_i \,,$ and $ \tilde \epsilon_i \equiv 2 m_\psi/m_i$, we can write the energy relation as follows. 
\begin{align}
	\frac{2 x_{n+1}}{1+\epsilon_{n+1} \eta_{n+1}} = x_n + |\beta_{n+1}| \sqrt{x_n^2 - \tilde \epsilon_n^2} \cos\theta_n \,.
	\label{eq:en_relation}
\end{align}
This relation constrains the form of the energy spectrum in the rest frame of $\phi_{n+1}$ obtained from that in the rest frame of $\phi_{n}$.

Next, we consider the relation of the energy spectra of the particle $\psi$ in the rest frames of $\phi_{n+1}$ and $\phi_{n}$.
The spectrum is a function of a dimensionless energy $x_{n+1}$ and the production angle $\theta_{n+1}$, $d N_\psi = p_{n+1} (x_{n+1}, \theta_{n+1}; \epsilon) d x_{n+1} d \cos\theta_{n+1}$. 
Here, $\epsilon$ collectively denotes dimensionless mass parameters.
Changing the variables to the parameters in the rest frame of $\phi_{n-1}$, one obtains
\begin{align}
	d N_\psi & = p_{n+1} (x_{n+1}, \theta_{n+1}; \epsilon) \nonumber \\
	& \quad \times |J(x_{n+1},\cos\theta_{n+1}|x_{n}, \cos\theta_{n})| d x_{n} d \cos\theta_{n} \nonumber \\
	& = p_{n} (x_{n}, \theta_{n}; \epsilon) d x_{n} d \cos\theta_{n} \,,
	\label{eq:pn_relation}
\end{align}
where $J$ denotes the corresponding Jacobian.
In the second line, we use the fact that the energy spectrum is also a function of $x_{n}$ and $\theta_{n}$ in the rest frame of $\phi_{n}$.
We obtain the relation between the spectra in two rest frames. 
The energy spectrum is given by the angular integral of the spectrum:
\begin{align}
	\frac{d N_\psi}{dx_{n+1}} = \int p_{n+1} (x_{n+1}, \theta_{n+1}; \epsilon) d \cos\theta_{n+1} \,.
\end{align}
This integral may be rewritten in terms of $x_{n}$ and $\theta_{n}$ by the use of \cref{eq:pn_relation}. 
When inserting a delta function $\delta(x_{n+1}'-x_{n+1})$ to this integral and substituting the variables, one obtains 
\begin{align}
	\frac{d N_\psi}{dx_{n+1}} 
	& = \int d \cos\theta_{n+1} dx'_{n+1} \delta(x_{n+1}'-x_{n+1}) \nonumber \\ 
	& \quad \times \frac{p_{n} (x_{n}, \theta_{n}; \epsilon)}{|J(x_{n+1},\cos\theta_{n+1}|x_{n}, \cos\theta_{n})|} \nonumber \\ 
	& = \int d \cos\theta_{n} dx_{n} \frac{|J(x'_{n+1},\cos\theta_{n+1}|x_{n}, \cos\theta_{n})|}{|J(x_{n+1},\cos\theta_{n+1}|x_{n}, \cos\theta_{n})|} \nonumber \\
	& \quad \times \delta(x_{n+1}'-x_{n+1}) p_{n} (x_{n}, \theta_{n}; \epsilon) \,.
\end{align}
Here, the ratio of the Jacobians is one. 
Since $x_{n+1}'$ is written as a function of $x_{n}$ and $\theta_{n}$ via \cref{eq:en_relation}, we obtain the relation between the energy spectra as follows.
\begin{align}
	& \frac{d N_\psi}{dx_{n+1}} 
	= \int d \cos\theta_{n} dx_{n} p_{n} (x_{n}, \theta_{n}; \epsilon) \nonumber \\ 
	& \qquad \times \delta\left[x_{n+1} - \frac{1+\epsilon_{n+1} \eta_{n+1}}{2} x_{n} \right. \nonumber\\
	& \qquad \left. - \frac{1+\epsilon_{n+1} \eta_{n+1}}{2} |\beta_{n+1}| \sqrt{x_{n}^2 - \tilde \epsilon_{n}^2} \cos\theta_{n} \right] \,.
\end{align}
We note that the function $p_n (x_n, \theta_n; \epsilon)$ can be decomposed into the product of its energy part and its angular distribution, $p_n (x_n, \theta_n; \epsilon) = g_n (x_n, \epsilon) f_n (\theta_n)$. 

Let us consider the cascade decay of the scalar particle $\phi_1 \to 2 \phi_0 \to 2 (\psi + X)$ and compute the energy spectrum of $\psi$ in the $\phi_1$ rest frame.
The dimensionless parameters are given by $\epsilon_1 = 2 m_0/m_1$ and $\eta_1 = 0$, and the velocity of $\phi_0$ in the $\phi_1$ rest frame is $|\beta_1| = \sqrt{1-\epsilon_1^2}$.
As for isotropic decay of the scalar particle, the angular distribution is $f_0 (\theta_0) = 1/2$, which is normalized as the integral over $\cos\theta_{0}$ to be one.
The resulting energy spectrum in the $\phi_1$ rest frame is 
\begin{align}
	\frac{d N_\psi}{dx_1} 
	& = \int^1_{-1} d \cos\theta_{0} \int^1_{\epsilon_0} dx_{0} \frac{d N_\psi}{dx_0} \nonumber \\
	& \times \delta\left[2 x_1 - x_{0} - \sqrt{(1-\epsilon_1^2)(x_{0}^2 - \tilde \epsilon_{0}^2)} \cos\theta_{0} \right] \,.
\end{align}
This is consistent with Ref.~\cite{Mardon:2009rc}.

Next, we consider cascade decay to be a three-body decay. 
We discuss the dark meson decay chain with the off-shell dark photon, which includes a pair of the SM fermions $f \bar f$ in the final state.
We assign each four-momentum as $\pi'(q) \to A'(p_1) + f (p_2) + \bar f (p_3)$.
We now compute the three-body phase space factor in the center-of-mass frame of $\pi'$, i.e. $q = (\sqrt{s}, \mathbf{0})$.
It is convenient to define the dimensionless quantities, 
\begin{align}
	\xi_i \equiv \frac{2 p_i \cdot q}{q^2} = \frac{2 E_i}{\sqrt{s}} \,, \quad (i = 1,2,3) \,, 
\end{align}
and the three-body phase space integral is simplified as follows. 
\begin{align}
	d \Pi_3 & = \frac{d^3 p_1}{(2\pi)^3 2 E_1} \frac{d^3 p_2}{(2\pi)^3 2 E_2} \frac{d^3 p_3}{(2\pi)^3 2 E_3} \nonumber \\ 
	& \quad \times (2 \pi)^4 \delta^4 (q - p_1-p_2 -p_3) \nonumber \\
	& = \frac{s d\xi_1 d\xi_2 }{64 \pi^3} \frac{d \Omega_1}{4 \pi} \frac{d \Omega_{12}}{4 \pi} \nonumber \\
	& \quad \times \delta \left(\cos\theta_{12} - \frac{E_3^2 - (|\mathbf{p}_1|^2 + |\mathbf{p}_2|^2 + m_3^2)}{2 |\mathbf{p}_1| |\mathbf{p}_2|} \right) \,.
\end{align}
Here, $d\Omega_1$ denotes the spherical integral associated with $d^3 p_1$, and $d\Omega_{12}$ denotes the spherical integral of relative angles between $\mathbf{p}_1$ and $\mathbf{p}_2$.

Assuming that the SM fermions are massless, one obtains the energy spectrum of the three-body decay $\pi'(q) \to A'(p_1) + f (p_2) + \bar f (p_3)$ as a function of $\xi_1$ and $\xi_2$ is given in the text, see \cref{eq:spectrum_3body}.
The function $f(t)$ is given by the integration over $\xi_1$ and $\xi_2$.
\begin{align}
	f(t) & \equiv \int_{\xi_{1:\mathrm{min}}}^{\xi_{1:\mathrm{min}}} d\xi_1 \int_{\xi_{2:\mathrm{min}}}^{\xi_{2:\mathrm{max}}} d \xi_2 \frac{1 + t^2 - \xi_1}{(1-\xi_1)^2} \\
	& \quad \times \left[ \xi_1^2 + 2 \xi_1 (\xi_2 -1) + 2 (\xi_2^2 -2 \xi_2 -t^2 +1) \right] \,. \nonumber
\end{align}
The resultant function is given in \cref{eq:fx_3body}.
The integration range for $\xi_1$ and $\xi_2$ is determined by the kinematics: $\mathbf{p}_1 = 0$ for the minimum of $\xi_1$ and $\mathbf{p}_1 = - (\mathbf{p}_2 + \mathbf{p}_3)$ for the maximum of $\xi_1$, while $\mathbf{p}_2 = 0$ for the minimum of $\xi_2$ and $\mathbf{p}_3 = 0$ for the maximum of $\xi_2$.
\begin{align}
	\xi_{2:\mathrm{min}} & = \frac12 \left(2 - \xi_1 - \sqrt{\xi_1^2 - 4 t^2} \right)\,, \nonumber \\
	\xi_{2:\mathrm{max}} & = \frac12 \left(2 - \xi_1 + \sqrt{\xi_1^2 - 4 t^2} \right)\,,\\
	\xi_{1:\mathrm{min}} & = 2 t \,, \nonumber \\
	\xi_{1:\mathrm{max}} & = 1 + t^2 \,.
\end{align}

\section{Equivalence Theorem}

One of the dark mesons is absorbed as the longitudinal mode of the dark photon in the chiral composite ADM. 
Some dark-neutral baryons decay into the longitudinal dark photon through the dark meson.
However, the dark-neutral baryons do not couple to the dark photon at the level of kinetic terms.
In this appendix, we demonstrate the equivalence of the decay rate of the dark nucleon into dark meson and the dark photon at high energy. 
We consider dark nucleon transition into another (off-shell) dark nucleon with emitting dark mesons in terms of baryon chiral perturbation theory instead of the decay induced by the portal operators (\ref{eq:Intermediate_Portal}) and (\ref{eq:Intermediate_Portal_Ch}). 
After the transition, the (off-shell) dark nucleon is converted into antineutrino through the portal operators. 
Under the flavor symmetry $SU(3)_L \times SU(3)_R$ (their transformation matrices denoted by $L$ and $R$), the dark baryon fields transform as $B' \to L B' L^\dag$ and $\overline B' \to R \overline B' R^\dag$ and the non-linear sigma field $U = \exp(2 i \Pi'/f_{\pi'})$ transforms as $U \to L U R^\dag$. 
The interaction Lagrangian up to $\mathcal{O}(p^2)$ terms of chiral perturbation between the baryon fields and the non-linear sigma field are 
\begin{align}
	\mathcal{L}
	& \supset 
	2 i \mathrm{Tr} \left[ B'^\dag \overline\sigma^{\mu} D_{\mu} B' + \overline B^\dag \overline \sigma^{\mu} D_{\mu} \overline B' \right] \nonumber \\
	& \quad + 2 \mathrm{Tr} \left[ m_{B} \overline{B}' U^\dag B' U 
	+ m_{B}^\ast U^\dag B'^\dag U \overline{B}'^\dag \right] \nonumber \\
	& \quad + 2 i \mathrm{Tr} \left[ x \,(D_{\mu} U) U^\dag B'^\dag \overline{\sigma}^{\mu} B' 
	+ \overline x \, (D_{\mu} U)^\dag U \overline{B}'^\dag \overline{\sigma}^{\mu} \overline{B}' \right] \nonumber \\
	& \quad + 2 i \, \mathrm{Tr} \left[ y B'^\dag \overline{\sigma}^{\mu} (D_{\mu} U) U^\dag B' 
	+ \overline y \overline{B}'^\dag \overline{\sigma}^{\mu} (D_{\mu} U)^\dag U \overline{B}' \right] \,.
\end{align}
Here, $\mathrm{Tr}$ denotes the trace of flavor indices.
The $U(1)_D$ charge of left-handed dark quarks is proportional to the Cartan subgroup $U(1)_3'$ of the vectorial $SU(3)_V$ corresponding to a generator $\lambda_3$, while that of right-handed dark quarks differs from the left-handed counterpart by a factor of $a$.
Hence, the covariant derivatives of the baryon fields is
$D_{\mu} B' = \partial_{\mu} B' - i e' A'_{\mu} [\lambda^{3}, B']$ and $D_{\mu} \overline B = \partial_{\mu} \overline B - i e' a A'_{\mu} [\lambda^{3}, \overline B]$, while that of the non-linear sigma field is $D_{\mu} U = \partial_{\mu} U -ie' A'_{\mu} \lambda^{3} U + i e' a A'_{\mu} U \lambda^{3}$.
Since $(D_\mu U)^\dag = - U^\dag (D_\mu U) U^\dag$, the interaction terms are real when coupling constants $x \,, y \,, \overline x \,,$ and $\overline y$ are real.
The relevant interactions among the dark-neutral baryons and dark photon/dark meson are given by 
\begin{align}
	\mathcal{L} & \supset
	\frac{e}{\sqrt3} (1-a) (x+y) \left(\Lambda'^\dag \overline \sigma^\mu \Sigma'_3 + \Sigma_3'^\dag  \sigma^\mu \Lambda' \right) A'_\mu \nonumber \\
	& - \frac{1}{\sqrt3} (1-a) (x+y) \left(\Lambda'^\dag \overline \sigma^\mu \Sigma'_3 + \Sigma_3'^\dag  \sigma^\mu \Lambda' \right) \frac{\partial_\mu \pi_3'}{f_{\pi'}} \nonumber \\
	& - \frac{e}{\sqrt3} (1-a) (\overline x+\overline y) \left(\overline \Lambda'^\dag \overline \sigma^\mu \overline \Sigma'_3 + \overline \Sigma_3'^\dag  \sigma^\mu \overline \Lambda' \right) A'_\mu \nonumber \\
	& + \frac{1}{\sqrt3} (1-a) (\overline x+\overline y) \left(\overline \Lambda'^\dag \overline \sigma^\mu \overline \Sigma'_3 + \overline \Sigma_3'^\dag  \sigma^\mu \overline \Lambda' \right) \frac{\partial_\mu \pi_3'}{f_{\pi'}} \,.
\end{align}

Now, we explicitly calculate two amplitudes for a process $\Sigma_3' (p) \to \Lambda'(k) + A'_\mu(q)$ and for a process $\Sigma_3' (p) \to \Lambda'(k) + \pi'_3(q)$, and then we take a high-energy limit. 
For the former process, the amplitude is 
\begin{align}
	& i \mathcal{M} (\Sigma'_3 \rightarrow \Lambda' + A'_{\mu} ) \nonumber \\
	& = i \frac{e' (1-a)}{2 \sqrt{3}} (x + y + \overline x + \overline y) \overline{u}(k) \gamma^{\mu} u(p) \epsilon^\ast_{L\mu}(q) \nonumber \\
	& \quad - i \frac{e' (1-a)}{2} (x + y - \overline x - \overline y) \overline{u}(k) \gamma^{\mu} \gamma^{5} u(p) \epsilon^\ast_{L\mu}(q) \,.
\end{align}
Here, $\epsilon_{L\mu}$ denotes the longitudinal polarization of dark photon, and $\epsilon_{L\mu}(q) \to q_\mu/{m_{A'}}$ at high energy.
On the other hand, for the latter process, the amplitude is given by
\begin{align}
	& i \mathcal{M} (\Sigma'_3 \rightarrow \Lambda' + \pi'_3) \nonumber \\
	& = \frac{1}{2 \sqrt{3}} (x + y + \overline x + \overline y) \overline{u}(k)  \frac{\not{\!p}\, - \!\not{\!k}}{f_{\pi'}} u(p) \nonumber \\ 
	& \quad - \frac{1}{2 \sqrt{3}} (x + y - \overline x - \overline y) \overline{u}(k) \frac{\not{\!p}\, - \!\not{\!k}}{f_{\pi'}} \gamma_{5} u(p) \,.
	\label{eq:decayamp_NGboson}
\end{align}
Reminding that the dark photon mass is $e' (1-a) f_{\pi'}$ in the chiral ADM model, one can find the decay rates agree with each other.

The dark baryon matrix elements are computed in this effective theory, and one can find the elements are written in terms of the coupling constants in the effective theory. 
Meanwhile, as we discussed in \cref{app:DR}, the matrix elements for different processes are related to each other by CGCs. 
We verify that the matrix elements for the dark nucleon transitions (with the off-shell dark baryon in the final state) satisfy the CGCs relations.
We define the decay amplitude corresponding to the matrix element $A^{(1)}$ in \cref{app:DR} by the decay $\Sigma'_3 (p) \to \Lambda'(k) + \pi'_3(q)$.
\begin{align}
	i \mathcal{M} (\Sigma'_3 \rightarrow \Lambda' + \pi'_3) 
	= \frac{1}{\sqrt{5}} i \mathcal{M}^{(1)} \,.
\end{align}
Here, the prefactor is from the CGCs. 
Comparing these with \cref{eq:decayamp_NGboson}, one finds
\begin{align}
	i \mathcal{M}^{(1)} & = \frac{1}{2} \sqrt{\frac{5}{3}} (x + y + \overline x + \overline y) \overline{u}(k) \frac{\not{\!p}\, - \!\not{\!k}}{f_{\pi'}} u(p) \nonumber \\
	& \quad - \frac{1}{2} \sqrt{\frac{5}{3}} (x + y - \overline x - \overline y) \overline{u}(k) \frac{\not{\!p}\, - \!\not{\!k}}{f_{\pi'}} \gamma_{5} u(p) \,.
\end{align}
We can compute the decay $\Sigma' (p) \to \Sigma'_3(k) + \pi'(q)$ in the effective theory, and its decay amplitude corresponds to the second matrix element $A^{(2)}$ in \cref{app:DR}.
\begin{align}
	& i \mathcal{M} (\Sigma' \rightarrow \Sigma'_3 + \pi') \nonumber \\
	& \quad = \frac{1}{2} (x - y - \overline x + \overline y) \overline{u}(k) \frac{\not{\!p}\, - \!\not{\!k}}{f_{\pi'}} u(p) \nonumber \\
	& \qquad - \frac{1}{2} (x - y + \overline x - \overline y - 2) \overline{u}(k) \frac{\not{\!p}\, - \!\not{\!k}}{f_{\pi'}} \gamma_{5} u(p) \nonumber \\
	& \quad = \frac{1}{\sqrt{3}} i \mathcal{M}^{(2)} \,.
\end{align}
Then, one finds 
\begin{align}
	i \mathcal{M}^{(2)} & = \frac{\sqrt{3}}{2} (x - y - \overline x + \overline y) \overline{u}(k) \frac{\not{\!p}\, - \!\not{\!k}}{f_{\pi'}} u(p) \nonumber \\
	& \quad - \frac{\sqrt{3}}{2} (x - y + \overline x - \overline y - 2) \overline{u}(k) \frac{\not{\!p}\, - \!\not{\!k}}{f_{\pi'}} \gamma_{5} u(p) \,.
\end{align}
Using these relations, we confirm the relations among the decay matrix elements. 
The matrix elements for other decay processes are 
\begin{align}
	& i \mathcal{M} (p' \rightarrow \Sigma_3' + \pi'_3) \nonumber \\
	& = \frac{1}{2} (x + \overline y) \overline{u}(k) \frac{\not{\!p}\, - \!\not{\!k}}{f_{\pi'}} u(p) \nonumber \\ 
	& \quad - \frac{1}{2} (x - \overline y - 1) \overline{u}(k) \frac{\not{\!p}\, - \!\not{\!k}}{f_{\pi'}} \gamma_{5} u(p) \nonumber \\
	& = \frac12 \sqrt{\frac35} i \mathcal{M}^{(1)} 
	+ \frac{1}{2\sqrt3} i \mathcal{M}^{(2)} \,, 
\end{align}
\begin{align}
	& i \mathcal{M} (p' \rightarrow \Lambda' + K') \nonumber \\
	& = \frac{1}{2} (x + \overline y) \overline{u}(k) \frac{\not{\!p}\, - \!\not{\!k}}{f_{\pi'}} u(p) \nonumber \\ 
	& \quad - \frac{1}{2} (x - \overline y - 1) \overline{u}(k) \frac{\not{\!p}\, - \!\not{\!k}}{f_{\pi'}} \gamma_{5} u(p) \nonumber \\
	& = - \frac{1}{2\sqrt5} i \mathcal{M}^{(1)} 
	+ \frac{1}{2} i \mathcal{M}^{(2)}  \,.
\end{align}
We conclude that the effective theory consisting of non-linear sigma field and baryon fields correctly reproduces the relations of the decay amplitudes expected from the CGCs. 

\bibliography{apssamp}

\end{document}